\documentclass{iucrjournals}

\usepackage{graphicx} 
\usepackage{multicol}
\usepackage{amsfonts, amsmath, amssymb, amsthm}
\usepackage{breqn}
\usepackage{comment}
\usepackage{enumitem}
\usepackage{algorithm,algorithmicx,algpseudocode}
\usepackage{tikz}
\usepackage{multicol}
\usepackage{subcaption}
\usepackage[normalem]{ulem}
\newcommand{\mat}[1]{\boldsymbol{#1}}

\newcommand{\MM}[1]{{\color{magenta} [\textbf{MM}: #1]}}

\title{Reducing acquisition time and radiation damage: data-driven subsampling for spectro-microscopy}

\author[a, d]{Maike Meier\IUCrCemaillink{m.meier@rug.nl}\IUCrOrcidlink{0000-0002-0822-3706}}%
\author[a,b]{Lorenzo Lazzarino\IUCrEmaillink{lorenzo.lazzarino@maths.ox.ac.uk}\IUCrOrcidlink{0009-0000-2889-8711}}%
\author[a]{Boris Shustin\IUCrCemaillink{boris.shustin@stfc.ac.uk}\IUCrOrcidlink{0000-0001-8013-9552}}%
\author[a]{Hussam Al Daas\IUCrEmaillink{hussam.al-daas@stfc.ac.uk}\IUCrOrcidlink{0000-0001-9355-4042}}
\author[c]{Paul Quinn\IUCrEmaillink{paul.quinn@stfc.ac.uk}\IUCrOrcidlink{0000-0002-7607-4271}}

\affil[a]{Scientific Computing Department, STFC, Rutherford Appleton Laboratory, Harwell Campus, Didcot, Oxfordshire, OX11 0QX, UK}
\affil[b]{Mathematical Institute, University of Oxford, Oxford, OX2 6GG, UK}
\affil[c]{Ada Lovelace Centre, STFC Scientific Computing, Rutherford Appleton Laboratory, Didcot, OX11 0QX,UK}
\affil[d]{Bernoulli Institute for Mathematics, Computer Science and Artificial Intelligence, University of Groningen, Groningen, 9747 AG, The Netherlands}

\begin{document} 
\maketitle 

\begin{synopsis}
Data-driven sparse acquisition in spectro-microscopy reduces acquisition time and radiation dose while preserving information.
\end{synopsis}

\begin{abstract}
Spectro-microscopy is an experimental technique which can be used to observe spatial variations in chemical state and changes in chemical state over time or under experimental conditions. As a result it has broad applications across areas such as energy materials, catalysis, environmental science  and  biological samples. However, the technique is often limited by factors such as long acquisition times and radiation damage. We present two measurement strategies that allow for significantly shorter experiment times and total doses applied. The strategies are based on taking only a small subset of all the measurements (e.g. sparse acquisition or subsampling), and then computationally reconstructing all unobserved measurements using mathematical techniques. The methods are data-driven, using spectral and spatial importance subsampling distributions to identify important measurements. As a result, taking as little as 4-6\% of the measurements is sufficient to capture the same information as in a conventional scan. 
\end{abstract}

\newpage


\section{Introduction}

X-ray Absorption Spectroscopy (XAS) is a powerful technique for investigating the chemical state or short-range bonding within a material. Spectro-microscopy combines XAS and microscopy to also understand the spatial variations of these properties in materials. In a typical spectro-microscopy experiment, the sample is rastered through the beam and the absorption or X-ray fluorescence (XRF) across the raster grid to form an image. A schematic overview of a spectro-microscopy experiment is shown in Figure 1a. This process is repeated for various energies above and below the absorption edge of the element of interest, see Figure 1b. The resulting stack of absorption or XRF images versus energy typically corresponds to $10$’s or $100$’s of thousands of individual spectra. The application of the technique to large regions or in-situ processes can be limited by long acquisition times and sample degradation due to high X-ray dosage. 

Here we set out to mitigate these limitations by reducing the number of measurements taken. When performing a spectro-microscopy experiment, the fact that there are only a few chemical states allows us to greatly reduce the data using techniques like PCA \cite{lerotic2004cluster,lerotic2005cluster}. Datasets are reduced to a few core spectral components and a map describing the mixture of those components. Knowing that there are only a few spectral components and the data can be reduced or compressed, our acquisition process can be designed to sample the data in different ways to exploit this. In designing an efficient acquisition scheme, it is necessary to incorporate (mechanically) practical rastering or continuous scanning.
In particular, we propose two novel sparse data acquisition schemes that are data-driven in nature. The schemes result in different strategies for selecting subsampled raster scans, see Figure 1c. Whereas every spatial row of pixels is measured in full raster scans, only a subset of the rows are measured in a subsampled raster scan. 

Townsend et al. \cite{townsend2022undersampling} proposed a sampling scheme, \textit{robust random raster sampling},  where the sampled rows are selected at random for each energy, while ensuring that every row of the specimen is sampled at least once. It is shown that with this approach, one can achieve a subsampling ratio of $10-20$\% routinely, without noticeably sacrificing the quality of the dataset. The subsampling ratio refers to the percentage of measurements that are taken compared to the total number of measurements for a normally sampled scan. The datasets are of the same dimension, yet only $15-20$\% of the entries will be measured. The rest of the entries are reconstructed with a matrix completion algorithm. The aforementioned matrix completion approach was further generalized to tensor completion methods in \cite{townsend2025alternatingsteepestdescentmethods}, allowing further reduction in the subsampling ratio to $10$\% without significant degradation of the dataset. 

The Robust random raster sampling approach is used as a baseline. We improve upon this by adopting a data-driven, adaptive approach to measurement selection, rather than relying on fully random sampling. This method intelligently chooses which spatial rows to sample based on the data itself, enabling a significant reduction in subsampling ratios—well beyond what has previously been achieved—without compromising the quality of the reconstructed dataset.
Our results show that high-quality reconstructions are possible even at subsampling ratios as low as $4-6$\%. Across all tested ratios, the data-driven techniques consistently outperform uniform random sampling.

Reducing the measurement time, while still addressing the statistical significance and signal-to-noise required across the energy range in XAS, is critical to broadening the impact of XAS spectro-microscopy. While the increased flux of new sources helps to accelerate measurements, subsampling can further accelerate these approaches and help to manage dose. 

\begin{figure}[h]
    \centering
    \includegraphics[width=0.7\linewidth]{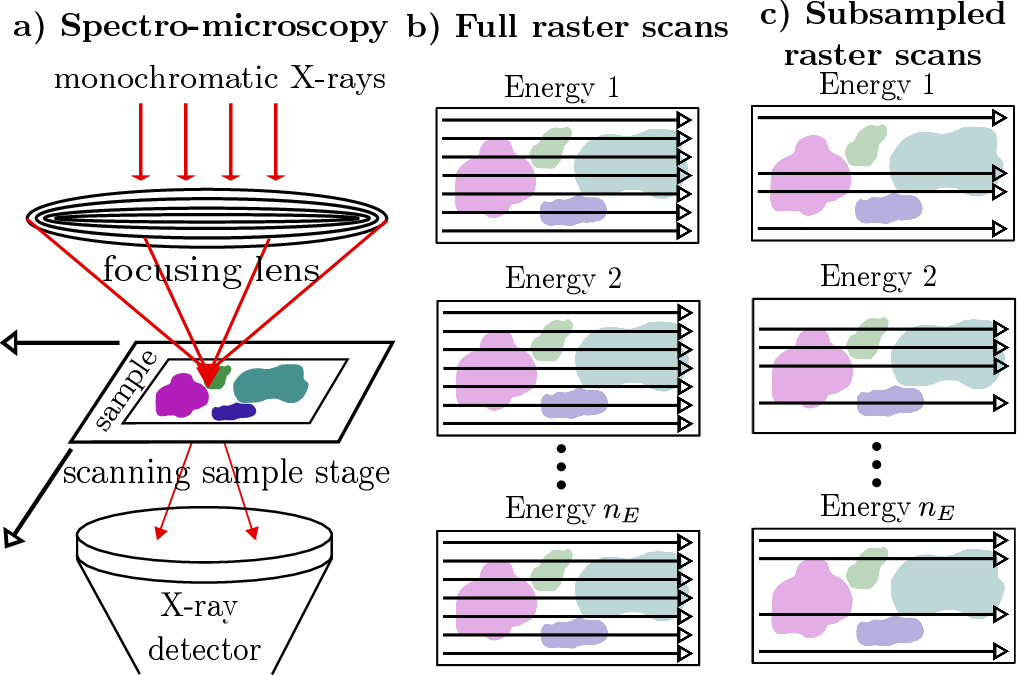}
    \caption{a) A schematic overview of a spectro-microscopy beamline (inspired by Hitchcock (2015). Monochromatic X-rays of a set energy are focused as the sample moves in a raster-like way. For full scans, the sample is scanned row-by-row as in b). One can reduce acquisition time and sample dose by only measuring a small subset of rows for each energy as in c).}
    \label{fig:introduction_rasters}
\end{figure}

\subsection{Spectro-microscopy}
In a spectro-microscopy experiment \cite{braun2005carbon,miller2017operando,hitchcock2005soft}, we have a sample that is exposed to X-rays of varying energy. We measure the absorption  $I(x,y,E)$ of the sample at position $(x,y)$ and at energy $E$, having fixed the incident flux $I_0(E)$. The optical density $D(x,y,E)$ is determined as $D(x,y,E) = -\ln\, [I(x,y,E)/I_0(E)]$ and is related to the absorption spectrum and thickness of the sample. Specifically, if the sample consists of one material with photon-energy-dependent linear absorption $\mu(E)$ at energy $E$, then the transmission image at $(x,y)$ is $I(x,y,E)=I_0(E)e^{-\mu(E)t(x,y)}$, where $t(x,y)$ is the thickness of the absorbing material. As a result, the ground truth optical density is linear in the thickness $t(x,y)$ and absorption $\mu(E)$. In particular, for the exact ground truth optical density $D_{\text{exact}}(x,y,E) = \mu(E)t(x,y)$.

If there are $S$ spectroscopically distinguishable components in the sample with absorption spectra $\mu_s(E)$ and thickness maps $t_s(x,y)$, $1\leq s\leq S$, then we have 
\begin{equation}
    \label{eq:D}
D_{\text{exact}}(x,y,E) = \sum_{s = 1}^S\mu_s(E)t_s(x,y).
\end{equation}

Mechanically, in a spectromicroscopic experiment, the incident beam is set to some energy and the sample is moved through the beam in a raster-like fashion, covering the sample either row-wise or column-wise. The spatial lay-out of the sample is divided in a grid of pixels, say $n_X \times n_Y$, for a total of $N=n_Xn_Y$ pixels and the transmission image is measured for each of the pixels. This procedure is repeated for various discretized energies $E_1, \dots, E_{n_E}$. Finally, after normalising with respect to the incident flux and computing the logarithm, we have $n_E$ datasets of size $n_X\times n_Y$ of the optical density. Throughout this work, we flatten these datasets to a single matrix $\mat{D}$ of size $n_E\times N$, as is typically done for analysis purposes, e.g., \cite{lerotic2004cluster,lerotic2005cluster}, where each row represents a vectorized XRF map and each column corresponds to the measured absorption spectrum for a single pixel. Assuming noiseless data, \eqref{eq:D} translates into a PCA-like factorisation $\mat{D}_{\mathrm{exact}}=\mat{M T}$, where $\mat{M}$ is the discretised spectra and $\mat{T}$ is the thickness map. In practice the data collected is noisy and we have $\mat{D} = \mat{M T} + \mat{G}$, with $\mat{G}$ representing the noise.
The fundamental goal of a spectromicroscopic experiment is to approximate $\mat{M}$ and $\mat{T}$ (or only $\mat{M}$) using the noisy dataset $\mat{D}$.

\subsection{Low-dimensionality of spectro-microscopy data and redundancy in measurements}
The flattened dataset $\mat{D}$ of the sample's optical density has underlying low-dimensionality. The low-dimensional structure of the spectro-microscopy data is crucial to both subsampling techniques as well as the standard computational data analysis routine.

The exact optical density is exactly low-rank; it can be represented as the product of the matrix of spectra and the matrix of (flattened) thickness maps:

\tikzset{every picture/.style={line width=0.75pt}} 
\begin{center}
\begin{tikzpicture}[x=0.75pt,y=0.75pt,yscale=-1,xscale=1]

\draw  [fill={rgb, 255:red, 93; green, 107; blue, 241 }  ,fill opacity=0.4 ] (120,19.67) -- (290,19.67) -- (290,59.33) -- (120,59.33) -- cycle ;
\draw  [fill={rgb, 255:red, 80; green, 227; blue, 194 }  ,fill opacity=1 ] (310.76,20.33) -- (328.43,20.33) -- (328.43,60) -- (310.76,60) -- cycle ;
\draw  [fill={rgb, 255:red, 241; green, 93; blue, 129 }  ,fill opacity=0.4 ] (333.12,33.42) -- (503.12,33.45) -- (503.11,51.12) -- (333.11,51.12) -- cycle ;

\draw (184,31.73) node [anchor=north west][inner sep=0.75pt]    {$\boldsymbol{D}_{\mathrm{exact}}$};
\draw (99.57,34.78) node [anchor=north west][inner sep=0.75pt]  [font=\small]  {$n_{E}$};
\draw (199.48,4.45) node [anchor=north west][inner sep=0.75pt]  [font=\small]  {$N$};
\draw (292.67,36.78) node [anchor=north west][inner sep=0.75pt]  [font=\small]  {$=$};
\draw (309.43,34.78) node [anchor=north west][inner sep=0.75pt]  [font=\small]  {$\boldsymbol{M}$};
\draw (412.1,36.78) node [anchor=north west][inner sep=0.75pt]  [font=\small]  {$\boldsymbol{T}$};
\draw (311.88,2.2) node [anchor=north west][inner sep=0.75pt]  [font=\small]  {$S$};
\end{tikzpicture}
\end{center}

The measured optical density $\mat{D}$ is naturally noisy, so it will not have exact low-rank structure. However, the directions in the data corresponding to the $S$ different materials remain dominant over the directions corresponding to noise -- leading to approximate low-dimensionality. 

This low-dimensionality means there is a large redundancy in the measurement. We demonstrate this using a low-rank decomposition technique called a \textit{CUR decomposition}. Assume for simplicity that we have a dataset $\mat{D}_{\mathrm{exact}}$ corresponding to a sample with $S=2$ spectroscopically different components. Select two energies and measure full raster scans for these. These data correspond to two rows of $\mat{D}_{\mathrm{exact}}$, which we will call $\mat{R}\in\mathbb{R}^{2\times N}$. Similarly, imagine we could measure the full absorption spectra for just two pixels, i.e. two columns of $\mat{D}_{\mathrm{exact}}$, and collect these data in $\mat{C}\in\mathbb{R}^{n_E\times 2}$. We can reconstruct all of $\mat{D}_{\mathrm{exact}}$ using just $\mat{C}$ and $\mat{R}$.

\begin{figure}
    \centering
    \includegraphics[width=0.5\linewidth]{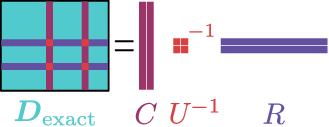}
    \caption{A matrix of exact rank $S$ (in this case $S=2$) can be represented using just $S$ of its columns and $S$ of its rows using the CUR decomposition. The only requirement is that the rows and columns are linearly independent. }
    \label{fig:CUR_intro}
\end{figure}

In particular, under only the assumption that the rows in $\mat{R}$ and the columns in $\mat{C}$ are linearly independent, we can form the CUR decomposition. Let $\mat{U}\in\mathbb{R}^{2\times 2}$ denote the intersection of $\mat{C}$ and $\mat{R}$, then $\mat{D}_{\mathrm{exact}} = \mat{C}\mat{U}^{-1}\mat{R}$. This is graphically displayed in Figure~\ref{fig:CUR_intro}. 

If we were able to take noiseless measurements, this means only $2(n_E + N)$ measurements rather than $n_EN$ measurements are required map the chemical state of a sample.

Our measurements are, however, naturally noisy, which means the dataset $\mat{D}$ is not exactly low-rank, but only approximately low-rank. As a result, any low-rank decomposition like the CUR decomposition will only approximate $\mat{D}$. Yet, it will still be true that $\mat{D} \approx \mat{C}\mat{U}^{-1}\mat{R}$ due to the low-dimensional nature. Practically, we will have to \textit{oversample} slightly: measuring a few more than two columns and rows to obtain a accurate approximation. We elaborate on this in Section~\ref{sec:CURintro}.

The CUR decomposition is just one approach to  illustrate the redundancy present in $\mat{D}$. In \cite{townsend2022undersampling}, the authors exploit the fact that there exists an approximate low-rank decomposition $\mat{D} \approx \mat{X}\mat{Y}$, where $\mat{X}\in\mathbb{R}^{n_E\times r}$ and $\mat{Y}\in\mathbb{R}^{r\times N}$ have one small dimension $r\ll n_E,N$. They suggest using subsampled raster scans chosen at random to reduce acquisition time, see Figure 1c, combined with a matrix completion algorithm, a looped alternating gradient-descent algorithm (LoopedASD), that tries to find such $\mat{X}$ and $\mat{Y}$. Generally speaking, the problem of matrix completion is to find the unknown entries of a matrix that is partially observed but has some underlying structure, such as low-dimensionality.

\begin{figure}
    \centering
    \includegraphics[width=0.5\linewidth]{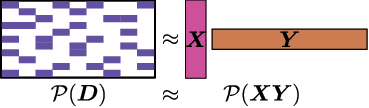}
    \caption{A graphical representation of random raster subsampling combined with the LoopedASD algorithm. Random raster sampling results in a dataset where the measurements correspond to blocks of entries for each row (left side). The LoopedASD algorithm aims to find two factors $\mat{X}$ and $\mat{Y}$ so that $\mat{XY}$ fits the observed measurements: $\mathcal{P}(\mat{D})\approx \mathcal{P}(\mat{XY})$.}
    \label{fig:low_rank_completion_intro}
\end{figure}

The process is illustrated in Figure~\ref{fig:low_rank_completion_intro}. By employing the subsampled raster scans, we observe just bits of the matrix $\mat{D}$. Let $\mathcal{P}(\mat{D})$ denote the matrix $\mat{D}$ with each unobserved entry set to zero. We then look for $\mat{X}$ and $\mat{Y}$ such that $\mathcal{P}(\mat{D}) \approx \mathcal{P}(\mat{X}\mat{Y})$; that is, $\mat{X}$ and $\mat{Y}$ are fitted only to the known entries. Although we have then only observed a subset of the measurements in $\mat{D}$, we hope that if $\mathcal{P}(\mat{D}) \approx \mathcal{P}(\mat{X}\mat{Y})$ it will be true that $\mat{D} \approx \mat{X}\mat{Y}$. Townsend \textit{et al.} (2022) use an alternating gradient descent-type algorithm, the so-called LoopedASD, to compute these matrices $\mat{X}$ and $\mat{Y}$. This method is described in detail in \cite{townsend2022undersampling}. 

Fundamentally, low-dimensionality not only allows us to compress data or analyse it with a PCA-like technique, but also to measure just a fraction of it and predict the rest reliably.

\subsection{The analysis of spectro-microscopic data}
\label{subsec:analysis-data}
In a perhaps more familar context, the low-dimensional nature is already exploited in the analysis of $\mat{D}$. Specifically, the analysis exploits the fact that the number of actual chemical states within the sample under study will be small and significantly less than the number of sampled positions \cite{lerotic2004cluster}. This allows the data to be described, to a high degree of accuracy, by a mixture of a small set of components.  

In particular, one first computes the eigenimages and eigenspectra corresponding to the data. Let $\mat{D}= \mat{Q}\mat{\Sigma}\mat{V}^T$ be a singular value decomposition of $\mat{D}$, where $\mat{Q}\in\mathbb{R}^{n_E\times n_E}$ and $\mat{V}\in\mathbb{R}^{N\times n_E}$ are orthonormal matrices and $\mat{\Sigma}$ is diagonal. The leading eigenimages are the first few rows of $\mat{\Sigma}\mat{V}^T$, and the corresponding eigenspectra are the leading columns of $\mat{Q}$.

Cluster analysis is performed on the eigenimages to cluster pixels that have similar eigenspectra. Finally, the average absorption spectra for a cluster can easily be found using the mean of the measured data for each cluster. We denote these spectra by $\mat{\hat{M}}_{\mathrm{cluster}}$. In the ideal case -- if the clusters correspond to a singular component -- these cluster spectra resemble the true spectra $\mat{M}$; if the clusters contain different components the cluster spectra will be linear combinations of the true spectra. The thickness maps $\mat{T}$ cannot reliably be reconstructed with this approach, because the natural solution $\mat{\hat{T}}_{\mathrm{cluster}} = \arg\min_{\mat{T}}\|\mat{D} - \mat{M}_{\mathrm{cluster}}\mat{T}\|_2$ is likely to have negative elements (which are unphysical).

\subsection{Subsampling spectro-microscopy measurements and data-driven approaches}
We have argued that it is possible to reduce the amount of measurements, while still recovering a good approximation to the full dataset $\mat{D}$. There  are two main questions that now arise: 1) what measurements should we take?, and 2) how do we then reconstruct the full data? The answer to the first question is restricted by the mechanical apparatus. Traditional fly-scanning approaches necessitate the scanning of full spatial sample rows (or columns), rather than individual pixels. As a result, the prior question boils down to a technique to select which spatial rows to sample.

In previous a work~\cite{townsend2022undersampling}, for each energy, sample rows were selected uniformly at random. The completion of the missing data was then done using LoopedASD. Another sampling approach, which is more closely related to our approaches, is presented in \cite{quinn2024optimal}. In the aforementioned work, energy samples are selected via a model order reduction technique, using a low-rank approximation of reference spectra or spectra measured in small parts of the specimen. Then, the full spectrum is recovered via the previously formed low-rank approximation. 

Recently, another line of works for subsampling the spectra uses methods of Baysian optimization. An AI-base approach of selecting energies to sample was presented in \cite{du2025demonstration}, where Bayesian optimization and Gaussian Process are the underlying techniques. In \cite{ito2025optimal}, Bayesian experimental design with Gaussian Process Regression is the underlying technique. 

We propose two new strategies for selecting which measurements to take, both are \textit{data-driven}. One of these strategies is based on the CUR decomposition \cite{drineas2008relative} and naturally allows for an optimization-free completion. The second strategy employs the same LoopedASD completion, although we note that any other matrix completion algorithm could also be used, as the sampling is independent of the matrix completion algorithm. 

\begin{figure}
    \centering
    \includegraphics[width=0.7\linewidth]{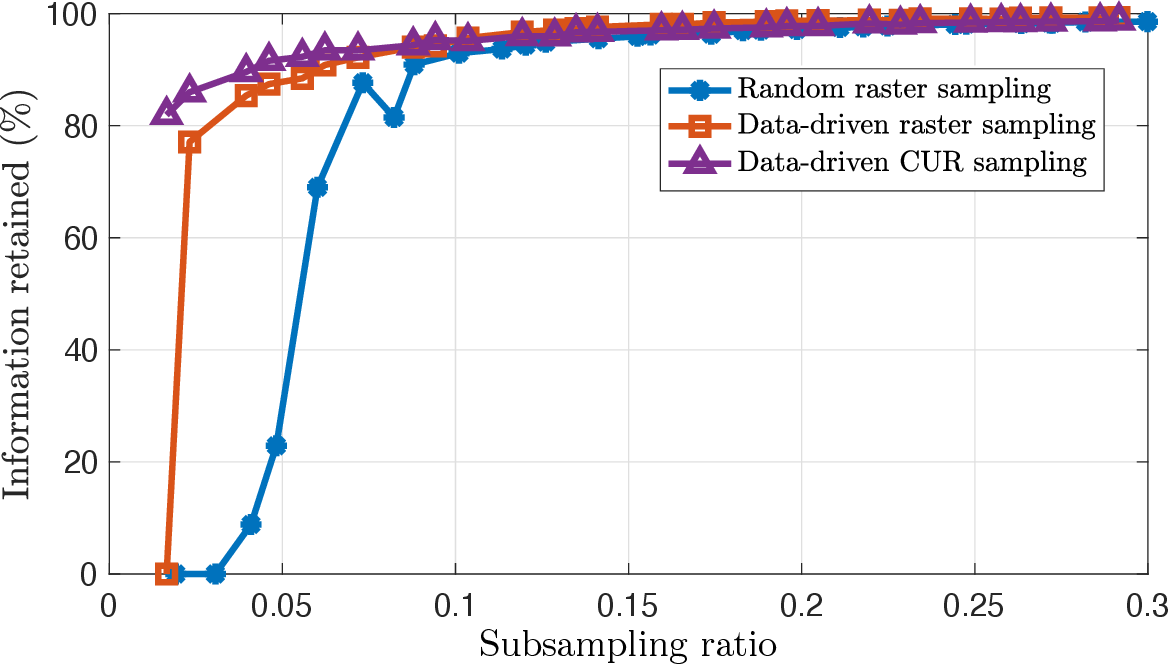}
    \caption{The amount of information retained with different subsampling strategies. The information measure is $(1 - \|\mat{D}-\mat{D}_{\mathrm{completed}}\|_F)/(1-\|\mat{D} - \mat{D}_{\mathrm{optimal}}\|_F)\cdot100\%$, where $\mat{D}_{\mathrm{completed}}$ is the completed dataset (with filled in entries) and $\mat{D}_{\mathrm{optimal}}$ is the best rank-$10$ approximation to $\mat{D}$ according to the Eckart-Young-Mirsky
Theorem \protect\cite{eckart1936approximation}, i.e., truncated rank-$10$ singular value decomposition (SVD). The dataset is DS1 from \protect\cite{townsend2022undersampling}. For the raster sampling approaches, completion is done with LoopedASD, and the data shown is the average of 10 iterations. For the CUR sampling approaches, the CUR decomposition is the completion and the data shown is the average of 100 iterations.}
    \label{fig:information_intro}
\end{figure}

We motivate our data-driven approaches over (uniform) random subsampling approaches in Figure~\ref{fig:information_intro} by comparing three different subsampling strategies. \textit{Random raster sampling} refers to the method in~\cite{townsend2022undersampling}, where for each energy a number of spatial rows is measured according to a uniform random distribution, and the matrix completion is done with LoopedASD. \textit{Data-driven raster sampling} is similar, but the spatial rows are selected based on an importance distribution on the sample (details on the importance distribution are in Subsection \ref{subsec:spatial-importance}). \textit{Data-driven CUR sampling} refers to taking measurements that correspond to full rows and columns in $\mat{D}$. That is, full XRF scans for some energies (rows) and measurements of some spatial rows for every energy (column blocks). As previously, the spatial rows are selected based on an importance distribution on the sample.

Figure~\ref{fig:information_intro} shows that we are able to retrieve a very large portion of the information accurately with the two new data-driven methods for very small subsampling ratios. The data-driven nature allows us to concentrate measurements where they are most useful, achieving subsampling ratios as small as $4\%$ while accurately retrieving $90\%$ of the information. Additionally, the figure shows that CUR completion is much more reliable than LoopedASD completion for very small subsampling ratios.

The methods we suggest are based on recent theory in randomized numerical linear algebra, where the CUR decomposition is a well-established tool. To approximate a general (approximately low-rank) matrix with a few number of its rows and columns, one needs to carefully select particular rows and columns that contain the most necessary information. The measures for the `information-richness' of rows or columns are called \emph{leverage scores} \cite{mahoney2011randomized}. Our data-driven methods are based on leverage-score-type schemes to ensure we are optimally exploiting every measurement. In Section~\ref{sec:impsubsampl} we explain the importance distributions, and in Section~\ref{sec:expapproaches} we suggest schemes for reducing spectro-microscopy acquisition times.

\subsection{Contributions}

In this subsection, we summarize our main contributions:

\begin{itemize}
    \item We propose two data-driven sampling approached for spectro-microscopy, aimed at reducing experiment time and dose applied on the specimens. Both strategies incorporate the use of statistical leverage scores to determine the importance distributions, both of the energy levels (when a spectral dictionary is available) and the spatial rows of the sample (Section \ref{sec:expapproaches}). 
    \begin{itemize}
        \item The first sampling strategy allows the inpainting of missing entries via any matrix completion algorithm such as LoopedASD (Subsection \ref{subsec:RISS}). 
        \item The second sampling strategy is designed specifically to allow the use of the CUR matrix factorization for approximating the missing entries (Subsection \ref{subsec: CURISS}).
    \end{itemize}
    \item We propose and explore adaptive approaches and stopping criteria for sampling (Subsection \ref{subsec:adaptivity-theory}).
\end{itemize}

\subsection{Outline}
We first discuss how to design data-driven importance subsampling distributions for the energy levels and spatial rows in Section~\ref{sec:impsubsampl}. The resulting subsampling and completion algorithms are presented in Section~\ref{sec:expapproaches}. In Section~\ref{sec:experiments} the algorithms are compared to the existing approach of random sampling.

\section{Spectral and spatial importance subsampling}\label{sec:impsubsampl}
We describe how spectral and spatial subsampling importance distributions can be formed. To this end we first introduce the general CUR decomposition.

\subsection{The CUR decomposition and columns/row subset selection}\label{sec:CURintro}
Let $\mat{A}\in\mathbb{R}^{m\times n}$ be a matrix with a small number of large singular values, and many small trailing singular values corresponding to noise. The rank-$k$ CUR decomposition \cite{goreinov1997theory,drineas2006fast} without oversampling of this matrix is defined by two index sets: $\mathcal{I}$ for the rows and $\mathcal{J}$ for the columns, both of which contain $k$ indices. We define the factors $\mat{C} = \mat{A}(:,\,\mathcal{J})$, $\mat{R} = \mat{A}(\mathcal{I},\, :)$, and $\mat{U} = \mat{A}(\mathcal{I},\mathcal{J})$, which are respectively the columns of $\mat{A}$ corresponding to $\mathcal{J}$, the rows of $\mat{A}$ corresponding to $\mathcal{I}$, and the intersection of these entries. The CUR approximation is given by $\mat{A}_{\mathrm{CUR}} = \mat{C}\mat{U}^{-1}\mat{R}\approx \mat{A}$. The CUR decomposition (without oversampling) exactly interpolates the measured entries, i.e. $\mat{A}(\mathcal{I},\,:) =  \mat{A}_{\mathrm{CUR}}(\mathcal{I},\,:)$ and likewise for the columns. All other rows and columns are approximated by linear combinations of the measured rows and columns, respectively.

We can also define a CUR decomposition where one of the index sets is oversampled \cite{drineas2008relative,mahoney2009cur}. Say $|\mathcal{I}| = k$ but $|\mathcal{J}| = k + p$ for some oversampling parameter $p$. We can then similarly define the rank-$k$ approximation $\mat{A}_{\mathrm{CUR}} = \mat{C}\mat{U}^{\dagger}\mat{R}\approx \mat{A}$, where $\mat{U}^{\dagger}$ indicates the pseudo-inverse of $\mat{U}$. This approximation will still interpolate the rows of $\mat{A}$ (i.e. $\mat{A}(\mathcal{I},\,:) =  \mat{A}_{\mathrm{CUR}}(\mathcal{I},\,:)$), but this will not generally hold for the columns. We are naturally restricted to the CUR decomposition with oversampling because in raster sampling whole blocks of columns are measured, rather than individual columns (pixels).

The quality of a CUR decomposition depends on the columns and rows that are chosen. That is, $\mathcal{I}$ and $\mathcal{J}$ determine the magnitude of the error $\|\mat{A}-\mat{A}_{\mathrm{CUR}}\|$. One particularly effective way of selecting good rows and columns is to use the associated leverage scores. The scores are a measure for the relative importance of a row or column. Focus on rows for simplicity, and let $l_i^2$ denote the leverage score of row $i$. The leverage scores are determined by the row-norms squared of an orthogonal basis for the dominant column space. That is, if $\mat{A} = \mat{U}_1\mat{\Sigma}_1\mat{V}_1^T +\mat{U}_2\mat{\Sigma}_2\mat{V}_2^T$ is an SVD of $\mat{A}$, where $\mat{U}_1\in\mathbb{R}^{m\times r}$ is the dominant column space, then the leverage scores for the rows of $\mat{D}$ are $l_i^2 = \|\mat{U}_1(i,\,:)\|_2^2$.

Now, the row leverage scores determine an excellent sampling distribution for the row index set $\mathcal{I}$. If we sample rows according to a probability distribution determined by $l_i^2$ (and use a similar process for the columns of $\mat{D}(\mathcal{I},:)$), then we have good theoretical guarantees on the quality of the CUR approximation. That is, an excellent strategy is to sample row $i$ with probability $l_i^2/n_E$.
Restricting the matrix $\mat{D}$ to the set of selected rows before determining the column indices is highly recommended for the CUR approximation, see \cite{park2025accuracy}.  

Exact leverage scores require access to the full matrix; however, in the case of a subsampled matrix as in this paper, estimations of the leverage scores are required. Various adaptive leverage score sampling strategies have been proposed in recent years. These strategies do not compute the leverage scores once and then sample from the resulting distribution, they update leverage scores based on what rows have already been sampled. One such algorithm is Adaptive Randomized Pivoting (ARP) see \cite{cortinovis2024adaptive}. Given an orthonormal basis of dimension $k$ for the columns space of $\mat{A}$, i.e. $\mat{U}_1$, the algorithm finds $k$ row indices that will result in small approximation errors for the CUR decomposition.

Of course, when it comes to our dataset $\mat{D}$, we do not have access to the leverage scores or a basis for the dominant row or columns space, because we have not yet observed the data. The main contribution of this work is a data-driven strategy that allows us to find good measurements to take without observing the dataset. 

\subsection{Determining a data-driven spectral importance distribution}
\label{subsec:spectral-importance}
We use the existing theory on the CUR decomposition and row subset selection problem to determine a \textit{spectral importance distribution}. This distribution is defined on the $n_E$ available energy levels, and should giver higher sampling probabilities to energies where more variance is explained. The leverage scores of $\mat{D}$ would be a great choice, but unfortunately we do not have access to those. However, an approximation is readily available if we have access to a dictionary of possible absorption spectra in the sample.

Considering the exactly low-rank (noiseless) matrix $\mat{D}_{\mathrm{exact}}$, which is close to $\mat{D}$. Since $\mat{D}_{\mathrm{exact}} = \mat{M}\mat{T}$, we can show that the leverage scores of $\mat{D}_{\mathrm{exact}}$ are the same as the leverage scores of $\mat{M}$, see Appendix \ref{appx:Motivation-dictionary}. If we have an approximation to $\mat{M}$, this means we can approximate the leverage scores of $\mat{D}_{\mathrm{exact}}$, and by that approximate the leverage scores of $\mat{D}$. This dictionary of absorption spectra $\mat{M}_{\mathrm{dictionary}}$ need not be exactly equal to $\mat{M}$ or even the same dimension as $\mat{M}$. 

\begin{figure}
    \centering
    \includegraphics[width=0.7\linewidth]{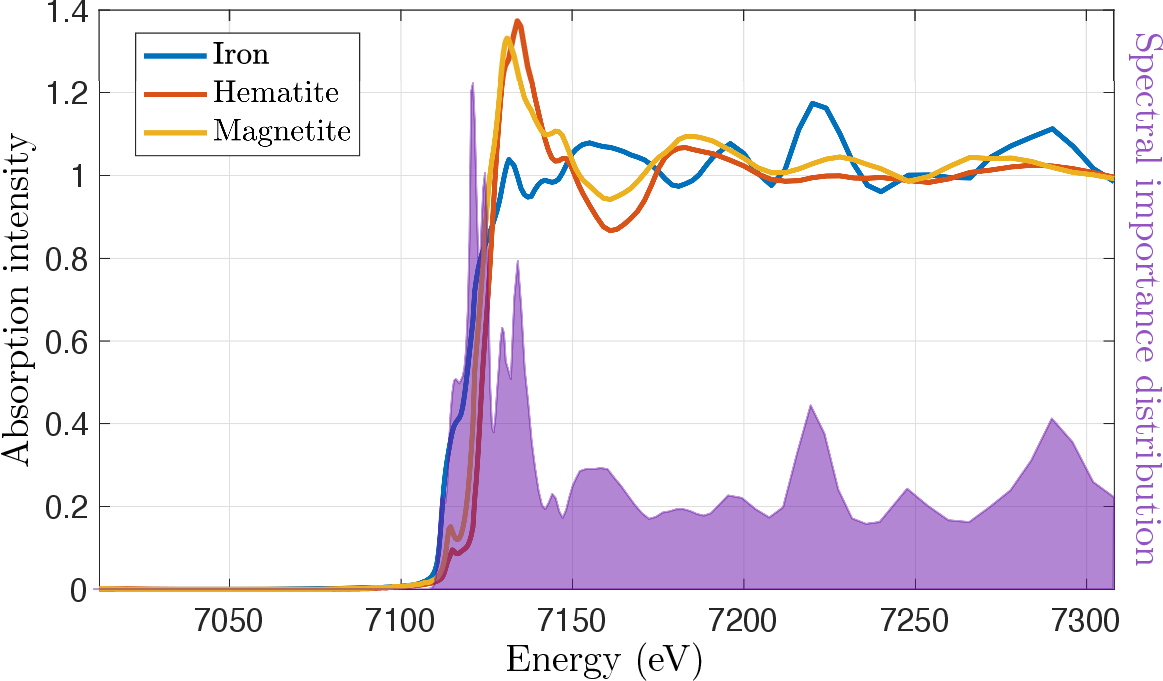}
    \caption{The spectral importance subsampling distribution (purple) for prior knowledge on the absorption spectra of three (potentially) present iron/iron-oxides. The incident energies (on the x-axis, in eV) where the most variance is explained are subsampled with higher probabilities. }
\label{fig:sampling_distribution_energy}
\end{figure}

We show an example distribution determined by leverage scores based on the dictionary spectra of iron, hematite, and magnetite in Figure~\ref{fig:sampling_distribution_energy}. One can interpret leverage scores as indicators for the the amount of variation in a dataset that is explained by a particular row. 

If a dictionary of true spectra $\mat{M}_{\mathrm{dictionary}}$ is not known, one may resort to random sampling a few energies. In Section~\ref{sec:noapriorispectralknowledge} we show how this adaptation affects the quality of the technique.

\subsection{Determining a data-driven spatial importance distribution}
\label{subsec:spatial-importance}
Selecting good columns of $\mat{D}$ to sample is a similar problem to energy selection. However, as we do not have access to a dictionary for the thickness maps, we need a different strategy. An additional mechanical limitation is that we can only measure entire spatial rows, not individual pixels (columns). A spatial row of the sample corresponds to a block of columns in $\mat{D}$. To  mitigate this we use a different unfolding of the data so that a row of the new dataset physically corresponds to a spatial row of the sample. This is then used to determine a good index set for the columns of $\mat{D}$ using an \textit{adaptive} leverage score sampling strategy. 

In the previous section we discussed how to determine an importance distribution on the energies. Assume we have selected a small number of energies, and that we have taken full raster scans at these energies. This will correspond to a few full rows of $\mat{D}$. These XRF scans provides us with an initial idea of the spatial lay-out of the sample, which will help us determine an approximate importance subsampling distribution. Denote the XRF scans with $\mat{F}_{s_1}, \dots, \mat{F}_{s_E}\in\mathbb{R}^{n_Y\times n_X}$, measured at respective energies $E_{s_1}, \dots, E_{s_E}$.

\begin{figure}
    \centering
    \includegraphics[width=0.7\linewidth]{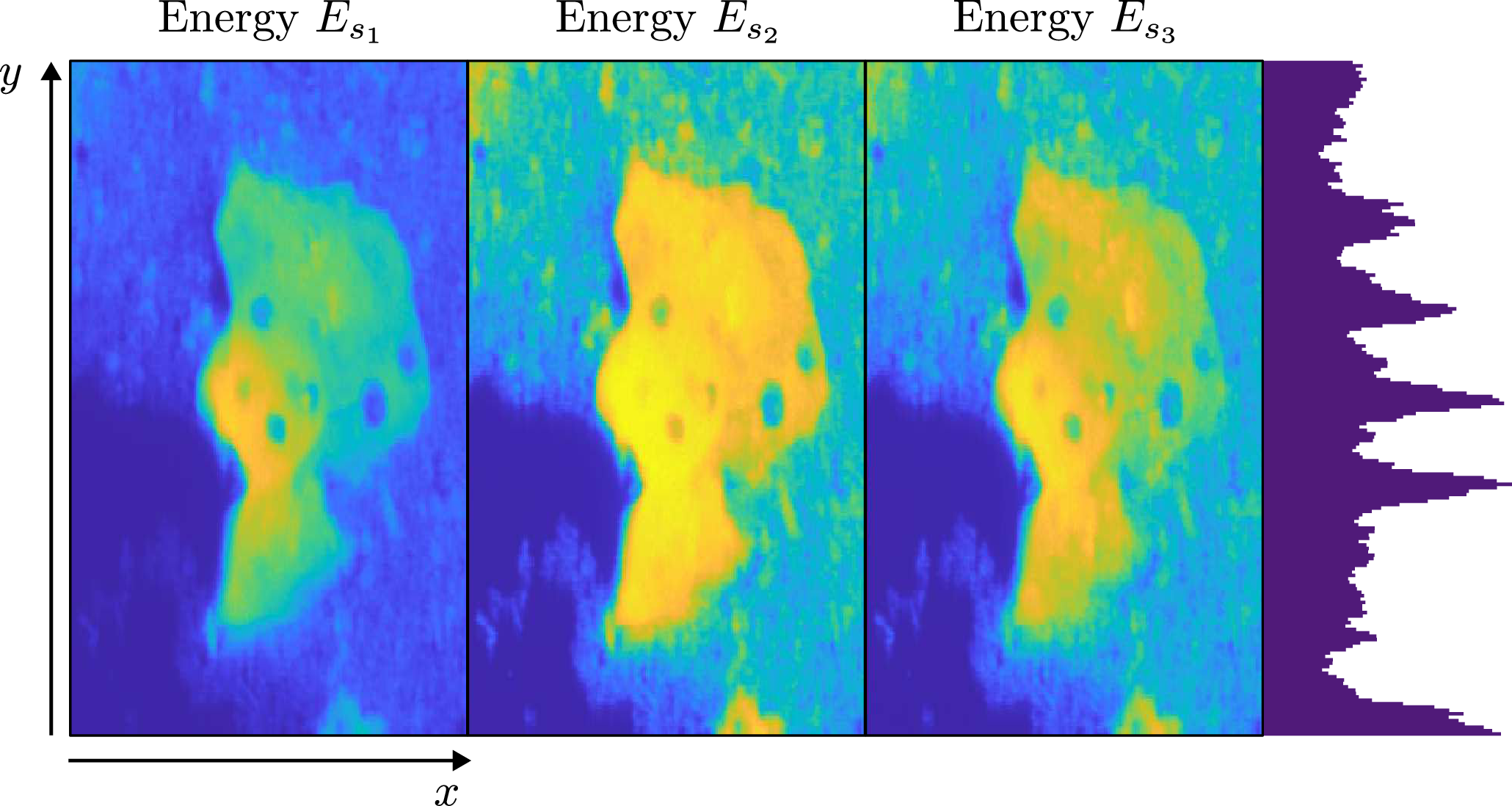}
    \caption{The importance subsampling distribution (purple) on the spatial rows. Three ($s_E = 3$) full raster scans at energies $E_{s_1}$, $E_{s_2}$, and $E_{s_3}$ are used to determine rows with high variance that are assigned a greater subsampling probability.}
\label{fig:sampling_distribution_spatial_rows}
\end{figure}

Next consider a different unfolding of the data. Define $\mat{F}\in\mathbb{R}^{n_Y\times s_En_X}$ as the matrix with individual XRF images stacked horizontally (see Figure~\ref{fig:sampling_distribution_spatial_rows} for an example). That is, $\mat{F} = \begin{bmatrix}
\mat{F}_{s_1} & \dots & \mat{F}_{s_E}
\end{bmatrix}.$ A row of $\mat{F}$ now corresponds to a spatial row of the sample. As a result, we can look at row selection for $\mat{F}$ to directly determine good sample positions. 

If we wish to sample $s_R$ spatial rows, we first compute the rank-$s_R$ truncated SVD of $\mat{F} = \mat{U}_1\mat{\Sigma}_1\mat{V}_1$. The dominant column space $\mat{U}_1\in\mathbb{R}^{n_Y\times s_R}$ is used to approximate the leverage scores associated to the spatial rows. We then use the adaptive leverage score sampling algorithm ARP to select $s_R$ spatial rows to sample from. The adaptive strategy allows us to assign smaller sampling probabilities to the areas of the sample that we have already sampled from, so increasing the probability to observe all important spatial areas.

Figure~\ref{fig:sampling_distribution_spatial_rows} shows an example distribution on the spatial rows using three XRF scans. The peaks of the distribution correspond to rows with high levels of variation in spectral make-up.

We have described methods to determine importance sampling distributions on the energies and spatial columns. In the next section we propose two different data-driven subsampling strategies for spectro-microscopy based on these ideas.


\section{Experimental approaches}\label{sec:expapproaches}
Armed with strategies for determining importance distributions on the energies and spatial rows, we propose two data-driven subsampling schemes. The first is a raster subsampling scheme similar to the scheme proposed in \cite{townsend2022undersampling}. The difference is that we use the importance distributions to sample from, rather than using uniform distributions. We call this approach Raster Importance Sampling for Spectro-microscopy (RISS). The second scheme is CUR-based, where the completion method is naturally the CUR approximation. We call this approach CUR Importance Sampling for spectro-microscopy (CURISS).

Both schemes start in the same way: 1) A spectral dictionary $\mat{M}_{\mathrm{dictionary}}$, if available, is used to determine a spectral importance distribution (as in Figure~\ref{fig:sampling_distribution_energy}), 2) sample a few energies from the spectral importance distribution and take full XRF scans at those energies, and 3) use these scans to determine a spatial importance distribution (as in Figure~\ref{fig:sampling_distribution_spatial_rows}).  The spectral dictionary can be formed from XAS standards, from fully-sampled scans taken previously during the experiment or a fusion of both. Otherwise, we can resort back to uniformly random sampling for the energies. We display the process in pseudocode in Appendix \ref{sec:appendix-algorithms}. In this algorithm we take the subsampling ratio as input, and then choose the number of sampled energies $s_E$ and the number of sampled spatial rows $s_R$ as approximately equal.

Note that small changes of temperature or mechanical movement of the equipment in the experiment, can lead to spatial drifts in the specimen during the experiment. When a full scan is available, this can be compensated during the stacking process via an alignment procedure using cross-correlation \cite{lerotic2014mantis}. In general, for subsampled scans, spatial registration to
correct drifts cannot be performed in the same way. However, in our method, full raster scans at selected energies are available, thus, various alignment methods can be still used to deal with drift. Though we do not implement it here, one can easily incorporate an alignment procedure into our method. Moreover, even if there is a small drift in the spatial rows, it is reasonable to assume that close by entries have similar importance, neglecting the effect of small drifts in our approach.

\subsection{RISS: Raster Importance Sampling for Spectro-microscopy} \label{subsec:RISS}
RISS is a data-driven variant of the raster subsampling method in \cite{townsend2022undersampling}. In this method, for each energy a small number of spatial rows is measured. These spatial rows are selected using a uniform random distribution at each energy. The adaptation we propose is to replace the uniform distribution with the spatial importance distribution. 

The measured dataset will have a small number of rows measured fully; these correspond to the full raster scans at the sampled energies. All other measurements will be column-wise blocks as in Figure~\ref{fig:low_rank_completion_intro}. These blocks will be concentrated around spatial rows with high sampling probabilities. The structure of the missing entries forces us to use a matrix completion algorithm. We choose the LoopedASD algorithm introduced in \cite{townsend2022undersampling}. See pseudocode in Appendix \ref{sec:appendix-algorithms}.

\subsection{CURISS : CUR Importance Sampling for spectro-microscopy} \label{subsec:CURISS}
CURISS is a CUR-based approach that makes use of the importance sampling strategies previously described and the CUR approximation as the completion method. The main steps are collected as in the beginning of this section.

Firstly, given a subsample ratio we calculate the number of full XRF scans and spatial rows that is possible to measure. 
Then, when possible, we determine the importance distribution of the XRF scans using the leverage scores associated to a dictionary of absorption spectra $\mat{M}_{\mathrm{dictionary}}$, as described in Section \ref{subsec:spectral-importance}. Alternatively, the spectral importance distribution is defined uniformily. We use these scores to select $s_E$ XRF scans to measure. These scans correspond to $s_E$ rows of $\mat{D}$. As explained in Section \ref{subsec:spatial-importance}, in order to determine spatial importance distribution we reshape the measured data into an $n_Y \times s_en_X$ matrix $\mat{F}$. We can now obtain importance scores for the rows of $\mat{F}$, corresponding to spatial rows of the sample using the ARP algorithm. 

The difference between RISS and CURISS is that we sample from this distribution just once, and measure these spatial rows for all remaining energies. As a result, the observed entries of the dataset $\mat{D}$ are full rows and blocks of columns. This allows us to employ the CUR decomposition as a tool for completion, rather than a gradient descent or other matrix completion algorithm. See pseudocode in Appendix \ref{sec:appendix-algorithms}. 

\subsubsection{Adaptive approaches and stopping criteria}
\label{subsec:adaptivity-theory}

In this subsection, we suggest a few approaches for adaptiveness and stopping criteria, which we explore empirically in Section \ref{subsec:adaptivity-experiments}. In Section \ref{sec:expapproaches}, we describe our proposed algorithms, where we assume a subsampling ratio is given a-priori, i.e., sampling budget, and it determines the stopping criteria for the number of full XRF scans and number of spatial rows to be sampled. Furthermore, the spectral importance distribution and the spatial importance distribution are set in advance as described in Section \ref{sec:expapproaches}, and do not change as in the previous subsections. 

However, the approaches above lead to the following two questions: how many measurements are needed to obtain a good enough completion, i.e., can the subsampling ratio depend on some stopping criteria, apart from the given sampling budget; how can we refine our sampling set and select full XRF scans and spatial rows in an adaptive manner. To address these questions, we propose the following Adaptive CURISS (ACURISS) approach: estimate the accuracy of a small CUR approximation, and, if needed, refine it. This process has two main components: the refinement steps, and the stopping criteria. The algorithm is detailed and explained in the following paragraphs.

First, we provide further details on adding new scans. Suppose that we have performed the CURISS method for an initial subsampling ratio, $p_0$, to obtain $\mat{\hat{D}}_{0}$. Then, we refine the approximation by including more XRF scans. To avoid a significant increase in the dead time needed to mechanically change energy, we propose to only add XRF scans while updating the CUR approximation after each new sample. That is, we only use the spatial rows already included in $\mat{\hat{D}}_0$. At each step, for a new XRF scan, we can use the available leverage scores of $\mat{M}_{\mathrm{dictionary}}$ and set to zero the scores of the already selected XRF scans, thus ensuring the selection of new energies. Afterwards, we update the CUR approximation with the newly scanned XRF to obtain $\mat{\hat{D}}_{1}$, and compute the corresponding subsampling ratio $p_1$. We repeat the process until some stopping criteria are satisfied. Finally, $\mat{\hat{D}}$ and $p$ are obtained. 

In addition, we propose two options for stopping criteria. After each refinement of the CUR approximation is done via the additional sample, we suggest using the change between refinements, inspired by the error metrics used for the experiments (Eq. \eqref{eq:completion-error} and \eqref{eq:spectral-error}), and using only the sampled entries. Note that in practice, we cannot compute Eq. \eqref{eq:completion-error} and \eqref{eq:spectral-error}, as they require having a full scan available. In addition, while the change in the completion error can be measured in the sampled entries, the spectral error requires access to the full matrix. Thus, we resort to comparing the completed matrix in the current refinement step, $\mat{\hat{D}}_{i}$, with the one obtained in the previous refinement step $\mat{\hat{D}}_{i-1}$. We define \textit{completion variation} as a stopping criteria relating to the completion error. For $i \geq 1$, stop when successive updates are sufficiently close:
\begin{equation}\label{eq:complettion-difference}
    \mathrm{Completion\, variation} = \|\mat{\Delta D}_{i}\| := \|\mat{\hat{D}}_{i} - \mat{\hat{D}}_{i-1}\|_F \leq \eta_{\mat{D}},
\end{equation}
where $\|\cdot\|_F$ denotes the Frobenius norm, and $\eta_{\mat{D}}\geq 0$ is a user-prescribed tolerance parameter. 

After each refinement step, we can perform a clustering analysis as described in Subsection \ref{subsec:analysis-data} and obtain approximations to the spectra, $\mat{M}_{\mathrm{cluster}}(\mat{\hat{D}}_{i})$. We compare these with those from the previous refinement step, $\mat{M}_{\mathrm{cluster}}(\mat{\hat{D}}_{i-1})$, and consider the magnitude of the update. The \textit{spectral variation} stopping criteria for $i \geq 1$ is met when
\begin{equation}\label{eq:spectral-difference}
    \mathrm{Spectral\, variation} = \|\mat{\Delta M}_{i}\| := \|\mat{M}_{\mathrm{cluster}}(\mat{\hat{D}}_{i}) - \mat{M}_{\mathrm{cluster}}(\mat{\hat{D}}_{i-1})\|_F  \leq \eta_{\mat{M}},
\end{equation}
where $\eta_{\mat{M}}\geq 0$ is again a tolerance parameter.

We remark that for both Eq. \eqref{eq:complettion-difference} and \eqref{eq:spectral-difference}, the average variation of several refinement steps can be used as alternative stopping criteria, rather than using a variation between only two refinment steps. As we demonstrate in Subsection \ref{subsec:adaptivity-experiments}, Eq. \eqref{eq:spectral-difference} gives us an indicator of the accuracy of the completion process. See pseudocode in Appendix \ref{sec:appendix-algorithms}.

\section{Experiments}\label{sec:experiments}
\subsection{Experimental Method}
The measurements were conducted on I14, the Hard X-ray Nanoprobe beamline at Diamond Light Source. The I14 endstation, high stability scanning stages, room and beam stability control measures are described elsewhere \cite{quinn2021hard, kelly2022delta, cacho2020passive,quinn2021beam} but in brief, the experiment was conducted in temperature controlled hutch using a nano-focussing KB mirror, with the sample rastered through the beam, in atmospheric conditions, using a voice coil actuated stage with interferometric control of the stage position, but no relative or differential control between the stage and KB.  The X-ray fluorescence signal (XRF) was collected using a 4-element silicon drift detector and energy changes and sparse scan motions were implemented using the Diamond acquisition software \cite{basham2018software}. 

The sample consisted of $\mathrm{Fe_2O_3}$ (hematite), and $\mathrm{Fe_3O_4}$ (magnetite) powders, ground, mixed and drop cast on a silicon nitride membrane to provide different mixtures and a well-known reference. While simulated spectra could have been used, experimental data capture realistic sample drift, beam artifacts such as top-up and other artifacts which allow us to test the approaches against real conditions. 
The XRF maps at each energy were collected in fly-scanning mode. The dwell time per point over the absorption edge was constant (15 ms).

Sample areas scanned ranged from $3\times 3$ $\mu m$ to $10 \times 10$ $\mu m$ using $50$, $75$ or $100$ $\mu m$ steps. The step size and area were chosen to keep the scanning time to a reasonable level during the time available for the overall experiment, rather than being a limit in resolution or size. 

We summarise the dimension of the datasets in Table~\ref{tab:dataset_summary}. Each of the datasets are obtained at I14 as described above.

\begin{table}[]
\centering
\begin{tabular}{lll|lll}
Name & $n_X\times n_Y$ & $n_E$ & Name & $n_X\times n_Y$ & $n_E$ \\ \hline
DS1  & $101\times 101$ & $149$ & DS5  & $80\times 80$   & $152$ \\
DS2  & $92\times 79$   & $150$ & DS6  & $76\times 79$   & $152$ \\
DS3  & $101\times 101$ & $149$ & DS7  & $86\times 73$   & $151$ \\
DS4  & $40\times 40$   & $152$ & DS8  & $85\times 84$   & $250$
\end{tabular}
\caption{The dimensions of each of the datasets used in Section~\ref{sec:experiments}. Datasets DS1-DS5 are identical to DS1-DS5 from Townsend \textit{et al.} (2022).}\label{tab:dataset_summary}
\end{table}

\subsection{Results}
We use two different measures to determine the quality of the the sampling strategies and the completion. The first is the distance between the completed dataset, say $\mat{\hat{D}}$, and the full dataset $\mat{D}$. We define
\begin{equation}\label{eq:completion-error}
    \mathrm{Completion\, error} = \frac{\|\mat{D} - \mat{\hat{D}}\|_F}{\|\mat{D}\|_F}.
\end{equation}

It is important to note that the size of this error is lower bounded by the error with respect to the optimal low-rank approximation \cite{eckart1936approximation}. Fundamentally, when approximating a matrix of full rank by a matrix that has significantly smaller rank, the error is controlled by the trailing singular values. As a result this error will likely not be less than, say, 0.01, because it will now correspond to 1\% noise present in the data. 

\begin{figure}
    \centering
    \includegraphics[width=\linewidth]{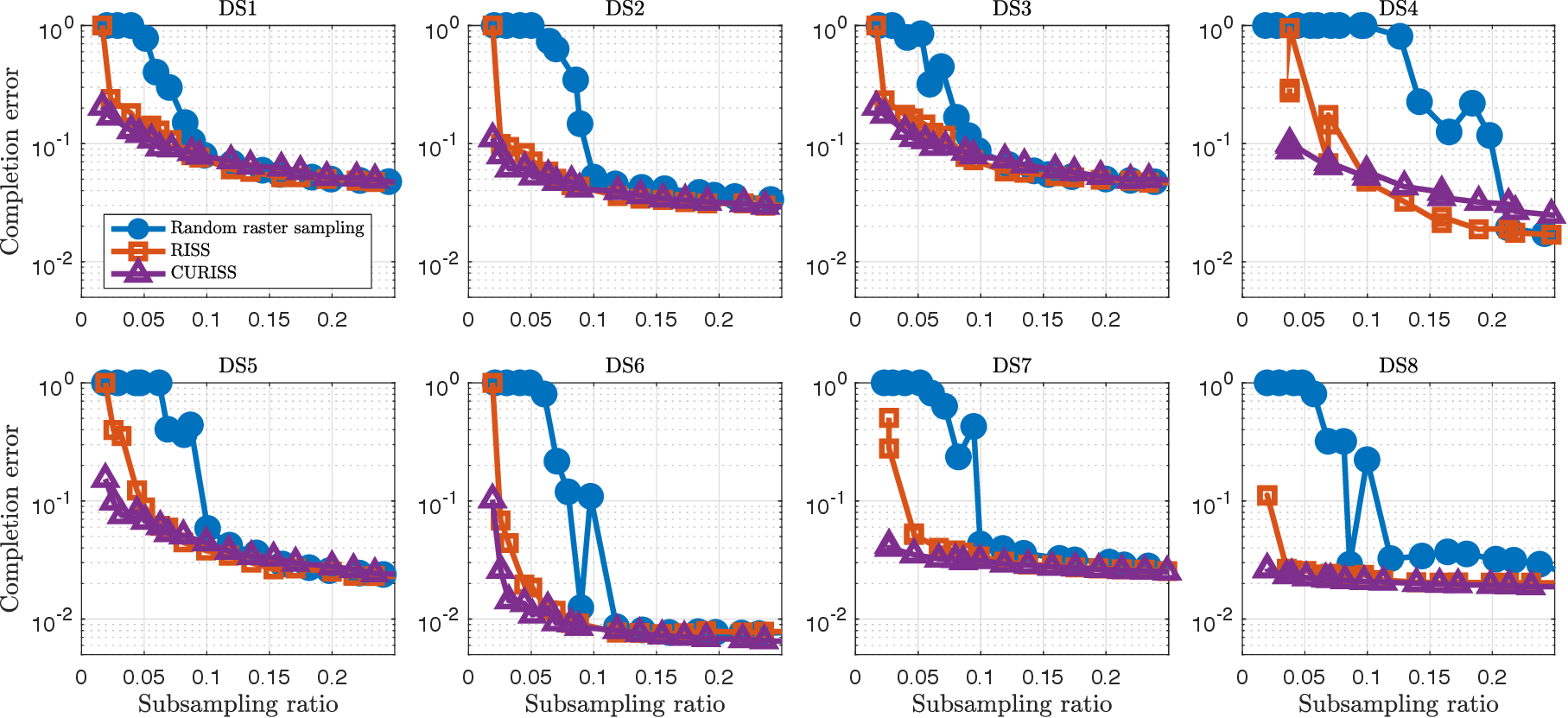}
    \caption{The completion error (defined in Eq.~\ref{eq:completion-error}) associated to each of the discussed algorithms: random raster sampling, data-driven raster sampling (RISS), and data-driven CUR raster sampling (CURISS). The data shown for random raster sampling and RISS is the average of 10 trials, wheras the data shown for CURISS is the average of 100 trials. LoopedASD is set to maximum rank 5 and tolerance $10^{-5}$. If LoopedASD fails, the error is set to equal $1$.}
    \label{fig:completion_error_main}
\end{figure}

Secondly, we define an error related to the further analysis. As described in Section~\ref{subsec:analysis-data}, the most common way to analyse spectro-microscopic data is to perform cluster analysis on the eigenimages and use average absorption spectra per cluster as estimates for the true absorption spectra $\mat{M}$. Denote the cluster spectra resulting from this analysis on the full dataset by $\mat{M}_{\mathrm{cluster}}(\mat{D})$ and on the completed dataset by $\mat{M}_{\mathrm{cluster}}(\mat{\hat{D}})$. We define
\begin{equation}\label{eq:spectral-error}
    \mathrm{Spectral \, error} = \frac{\|\mat{M}_{\mathrm{cluster}}(\mat{D}) - \mat{M}_{\mathrm{cluster}}(\mat{\hat{D}})\|_F}{\|\mat{M}_{\mathrm{cluster}}(\mat{D})\|_F}.
\end{equation}
The completion errors for eight different datasets are shown in Figure~\ref{fig:completion_error_main} (note the log-scale). We observe that RISS and CURISS generally perform well for subsampling ratios as small as $5\%$, where random raster sampling is generally unable to provide a satisfactory completion at this subsampling ratio. In most cases, CURISS outperforms RISS for subsampling ratios smaller than approximately $8\%$, after which RISS is the most succesful technique.

\begin{figure}[t]
    \centering
    \includegraphics[width=\linewidth]{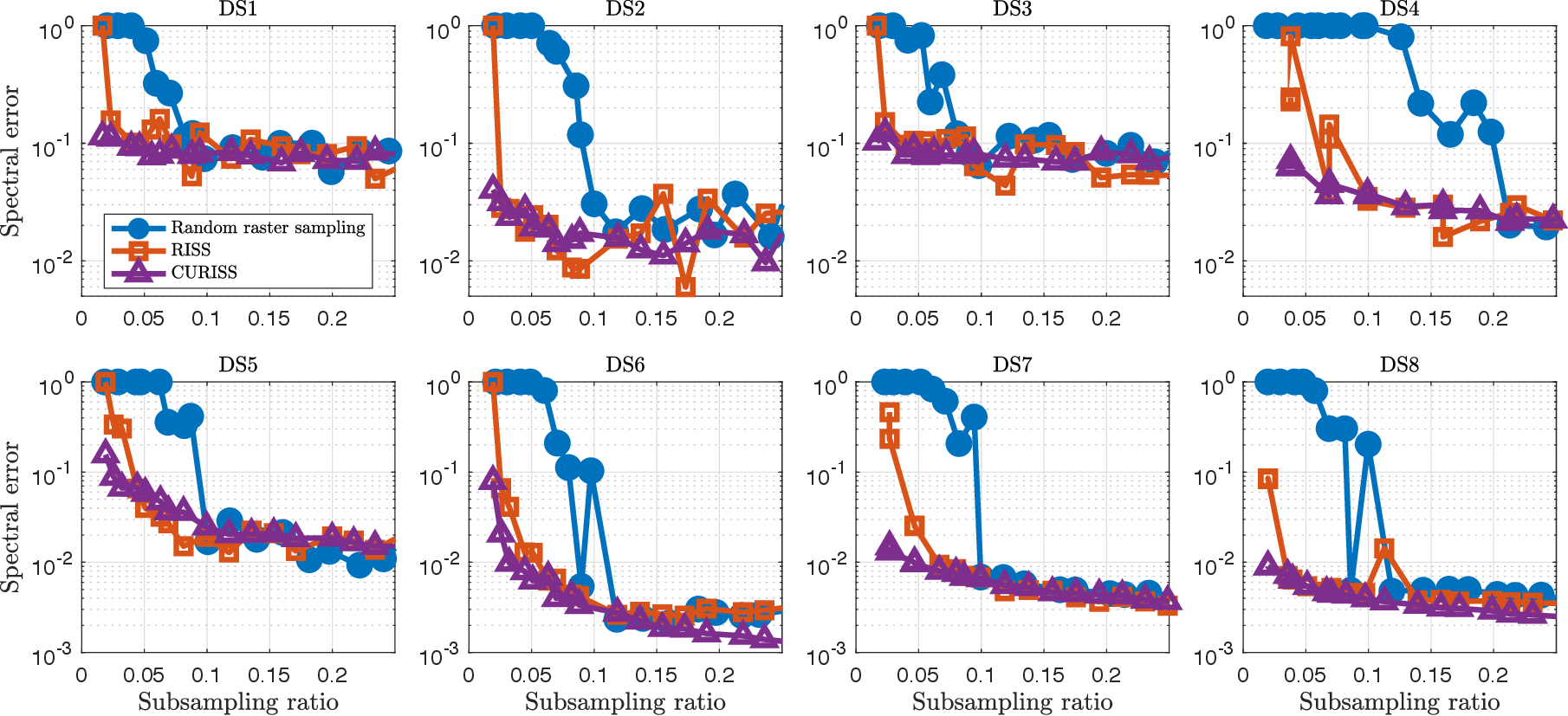}
    \caption{The spectral error (defined in Eq.~\ref{eq:spectral-error}) associated to each of the discussed algorithms: random raster sampling, data-driven raster sampling (RISS), and data-driven CUR raster sampling (CURISS). The experimental set-up is the same as in Figure~\ref{fig:completion_error_main}. We provide a close-up of the results for subsampling ratios less than $10\%$ in Figure~\ref{fig:spectral_error_zoom}.}
    \label{fig:spectral_error_main}
\end{figure}

\begin{figure}[H]
    \centering
    \includegraphics[width=\linewidth]{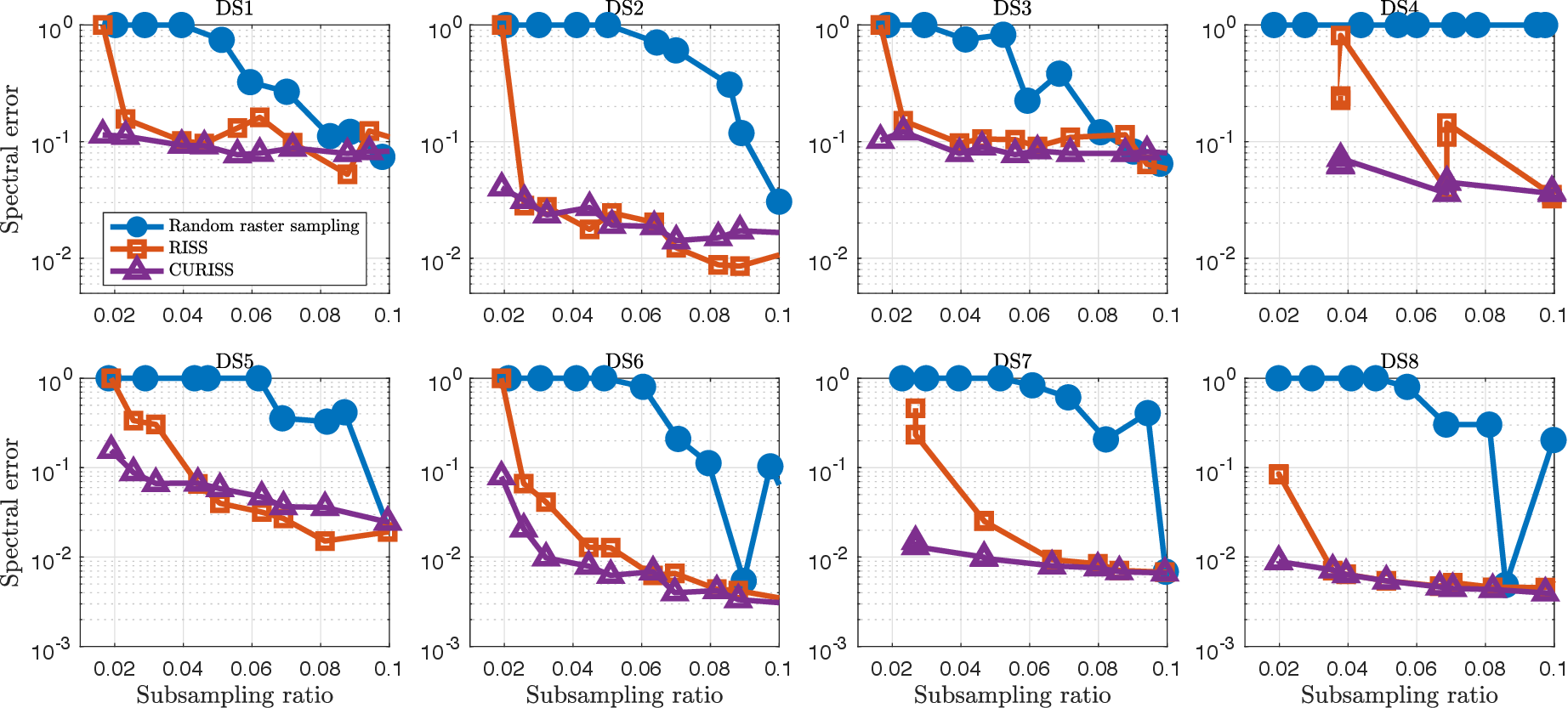}
    \caption{A close-up of Figure~\ref{fig:spectral_error_main} to observe subsampling ratios smaller than 10\%.}
\label{fig:spectral_error_zoom}
\end{figure}

The spectral errors for the same datasets are displayed in Figure~\ref{fig:spectral_error_main}, with a close-up of small subsampling ratios provided in Figure~\ref{fig:spectral_error_zoom}. It is even more evident in these figures that CURISS leads to good results for ratios as small as $3\%$. The CUR completion apparently captures the leading singular vectors and values (which are used for the cluster analysis) very well. RISS outperforms random raster sampling by at least an order of magnitude for these small subsampling ratios. 

We note furthermore that either of the methods that employ LoopedASD (random raster sampling and RISS) take considerably more computational effort, due to the optimization steps that they require. 
\subsubsection{Adaptive approaches}
\label{subsec:adaptivity-experiments}

\begin{figure}[t]
    \centering
    \begin{subfigure}[b]{0.24\textwidth}
        \centering
        \includegraphics[width=\textwidth, scale=0.5]{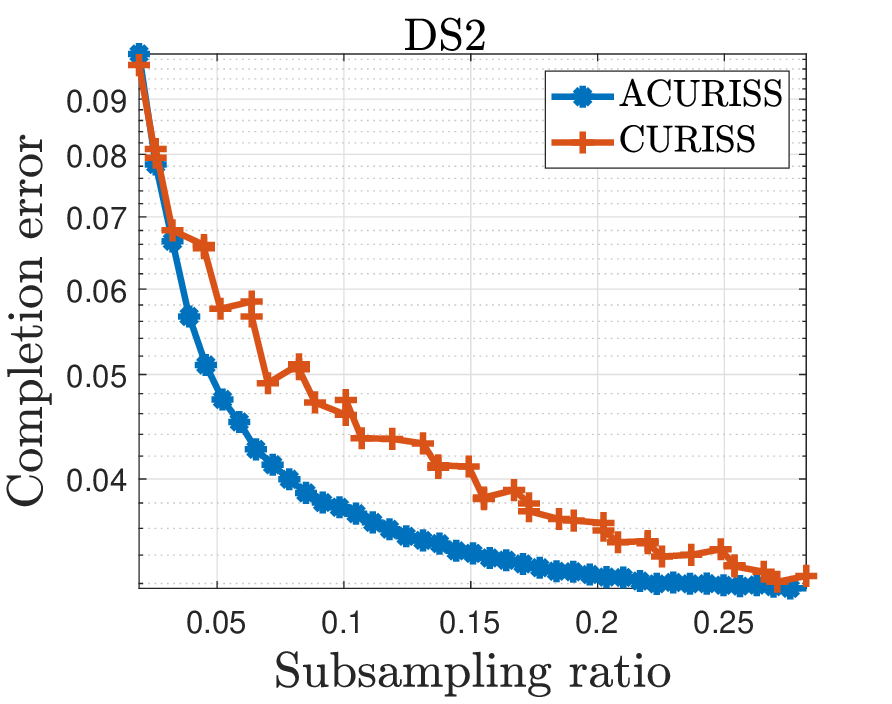}
    \end{subfigure}
    \begin{subfigure}[b]{0.24\textwidth}
        \centering
        \includegraphics[width=\textwidth, scale=0.4]{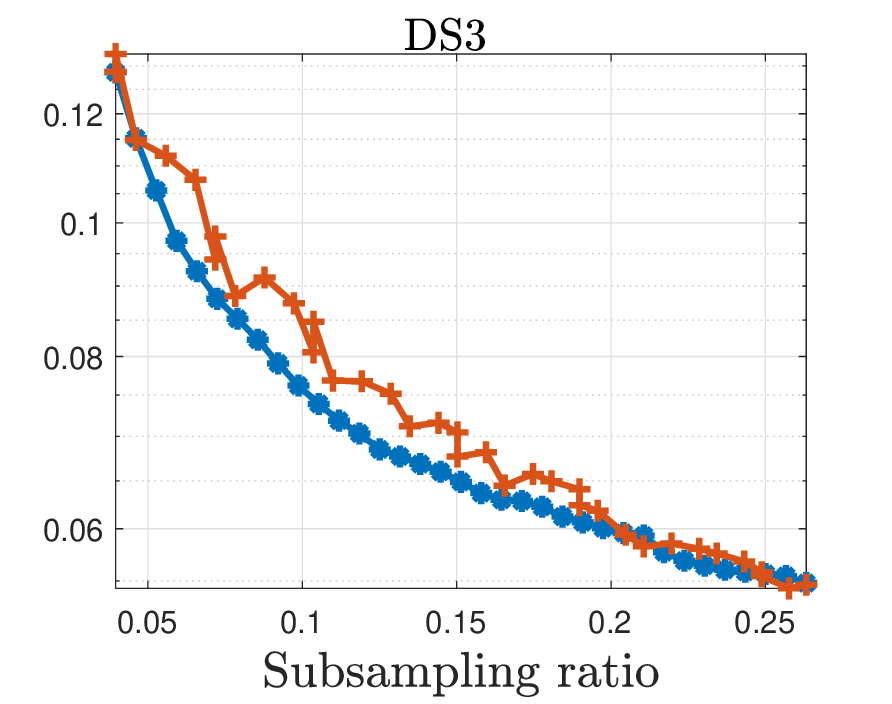}
    \end{subfigure}
    \begin{subfigure}[b]{0.24\textwidth}
        \centering
        \includegraphics[width=\textwidth, scale=0.4]{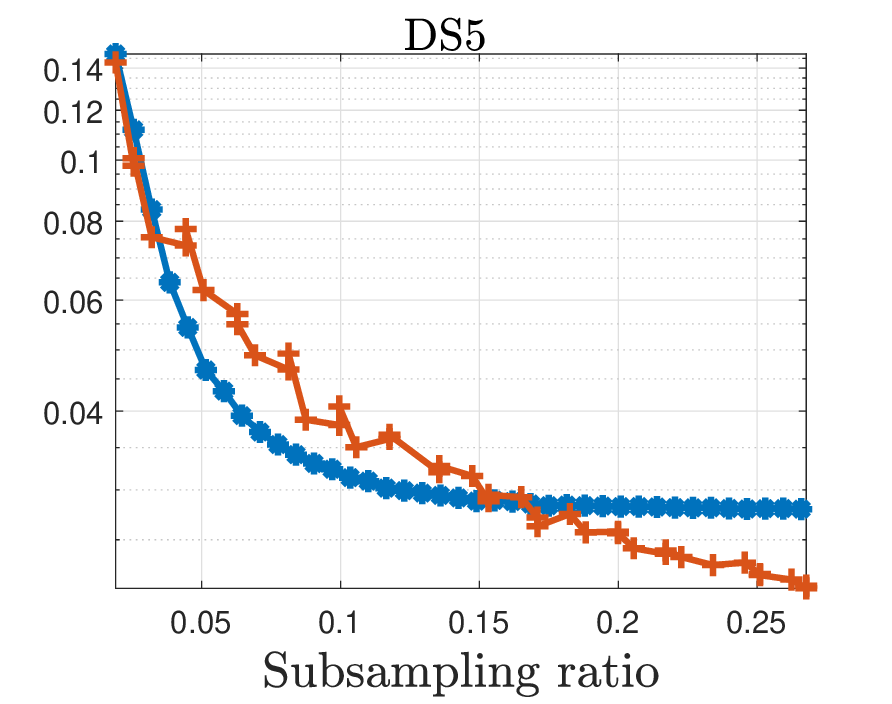}
    \end{subfigure}
    \begin{subfigure}[b]{0.24\textwidth}
        \centering
        \includegraphics[width=\textwidth, scale=0.4]{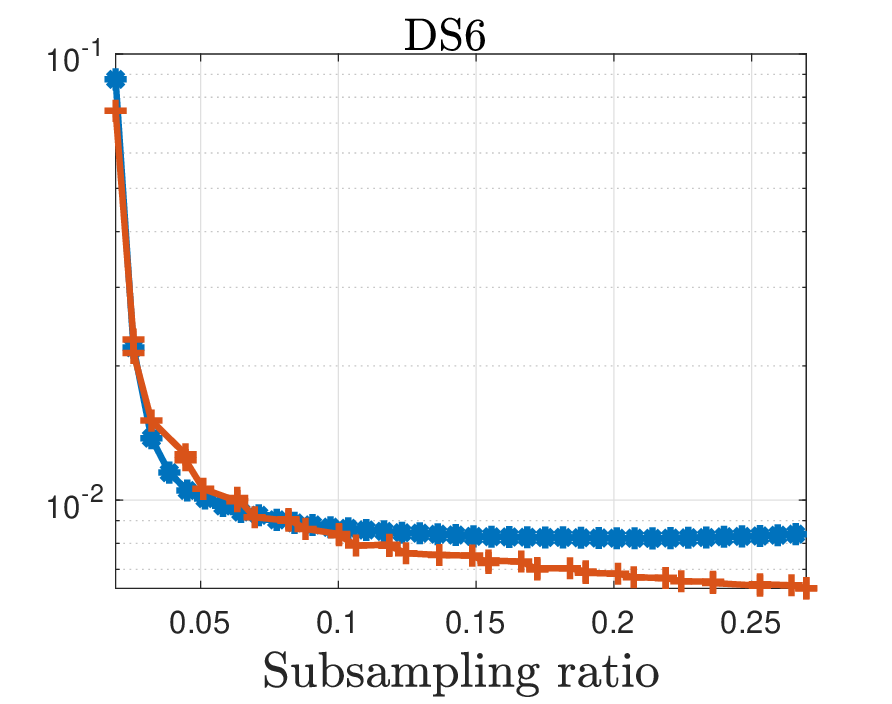}
    \end{subfigure}
    \caption{Comparison of completion errors for CURISS (orange) and ACURISS (blue). Displayed is the the mean error of $100$ trials. The shaded region indicates one standard deviation.}
    \label{fig:adaptiveVsCUR}
\end{figure}

\begin{figure}[t]
\centering
      \begin{subfigure}[b]{0.24\textwidth}
        \centering
        \includegraphics[width=\textwidth, scale=0.5]{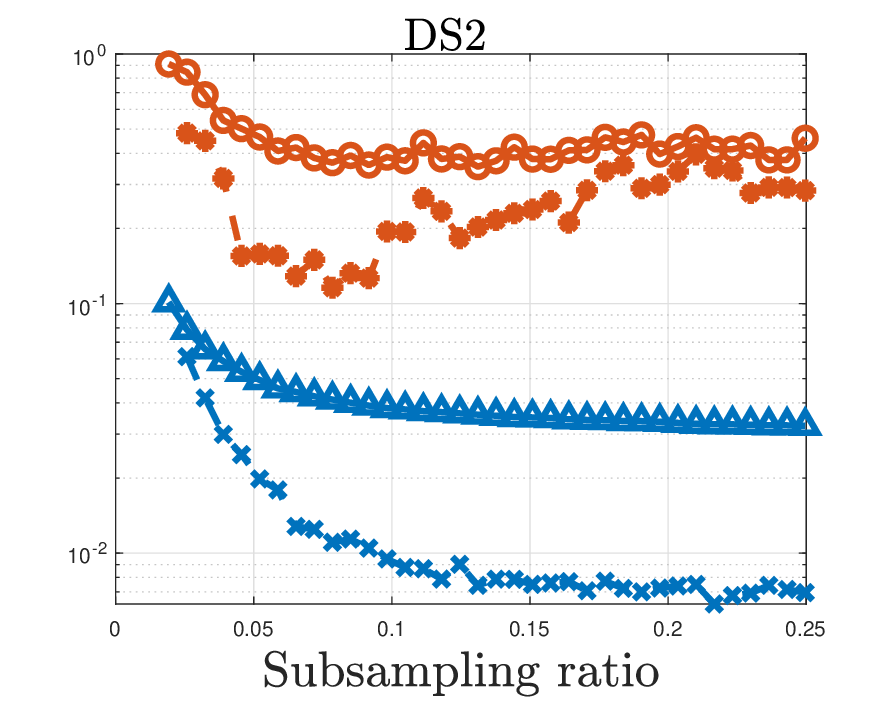}
    \end{subfigure}
    \begin{subfigure}[b]{0.24\textwidth}
        \centering
        \includegraphics[width=\textwidth, scale=0.4]{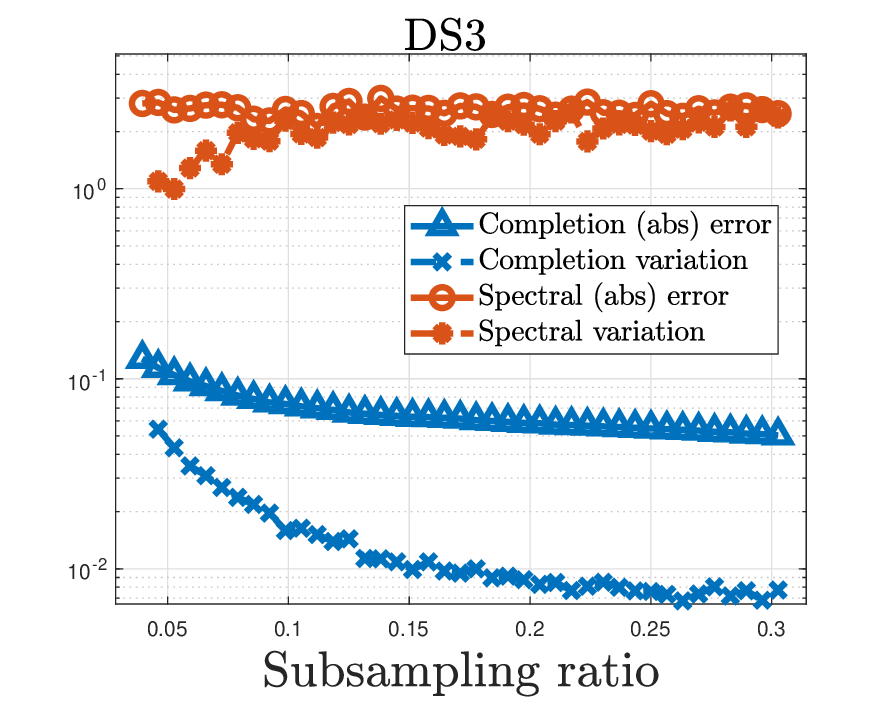}
    \end{subfigure}
    \begin{subfigure}[b]{0.24\textwidth}
        \centering
        \includegraphics[width=\textwidth, scale=0.4]{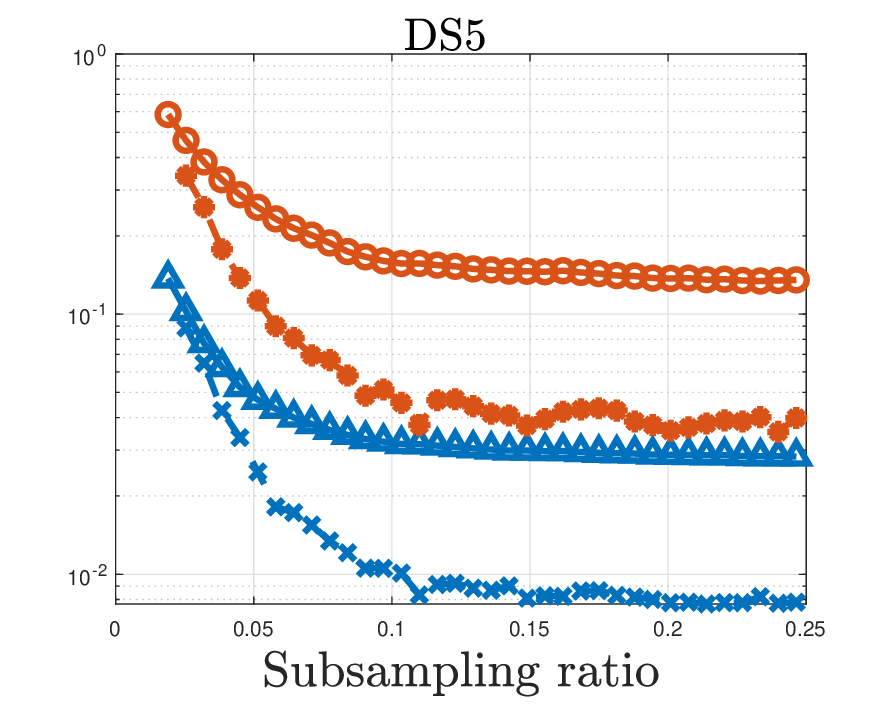}
    \end{subfigure}
    \begin{subfigure}[b]{0.24\textwidth}
        \centering
        \includegraphics[width=\textwidth, scale=0.4]{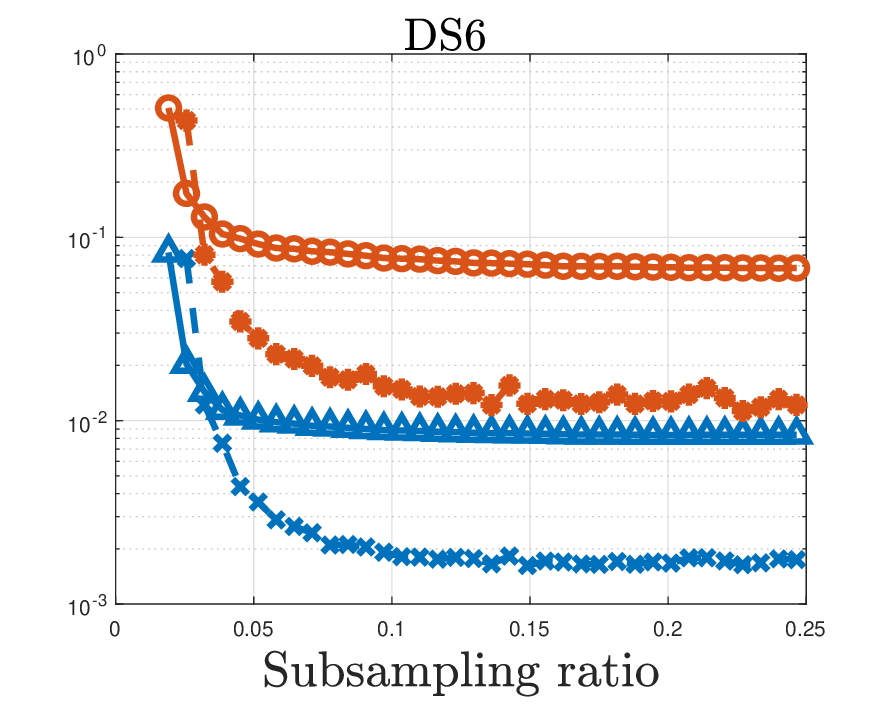}
    \end{subfigure}
\caption{Study of the proposed stopping criteria for the ACURISS strategy. After $100$ trials, the means of the completion absolute error for ACURISS (blue, solid line), the completion stopping criteria (blue, dashed line), the spectra approximation absolute error computed from the full matrix (orange, solid line), and the spectral stopping criteria (orange, dashed line), and  are presented.}
    \label{fig:adaptiveStopping}
\end{figure}
As described in Section \ref{subsec:adaptivity-theory}, the CURISS method allows for an adaptive counterpart that, starting with a small subsampling ratio, refines the CUR approximation by including new XRF scans as described in Subsection \ref{subsec:adaptivity-theory}. We perform experiments on different datasets to compare the accuracy—measured in completion error—of CURISS vs ACURISS. We begin by running ACURISS for a fix number of iterations. At each step, we store the used subsampling budget, $p_{i}$, and use it to compute the corresponding CURISS completed matrix. The choice of the initial subsampling ratio $p_0$ is crucial for the success of the adaptive strategy. We have observed that for most of the analyzed datasets $p_0 = 0.02$ is good enough. However, for other datasets a larger $p_0$ is needed to obtain completion errors competitive with the one obtained by CURISS. Specifically, DS3 needs $p_0 = 0.04$, while $p_0=0.1$ is necessary for DS8. 
Figure \ref{fig:adaptiveVsCUR} shows, for some datasets, that, with suitable $p_0$, the completion errors of the two strategies are comparable. This motivates further investigation of this technique.

As mentioned, an adaptive strategy requires a stopping criteria. Thus, we analyze how the two proposed techniques perform on different datasets. Since the considered variations \eqref{eq:complettion-difference} and \eqref{eq:spectral-difference} are absolute measures, we compare them with the respective absolute errors. Specifically, we compute refined approximations via the ACURISS and measure their completion absolute error. We then compute the completion variation (Eq. \eqref{eq:complettion-difference}) and the spectral variation (Eq. \eqref{eq:spectral-difference}). For analysis purposes, we also compute the spectral approximations attainable with the full matrix $\mat{D}$. In Figure \ref{fig:adaptiveStopping}, we present the main scenarios that may occur. In the best cases, the spectral variation is really close to the spectral error, e.g., for DS2 and DS3. Otherwise, the variations exhibit similar behavior to the actual errors but at different magnitudes. In both cases, it is clear that both stopping criteria can be used as indicators of the correspondent true errors. We note that the computational effort involved with updating the CUR decomposition and computing the stopping criteria is negligible. A similar scheme using LoopedASD would be more challenging, because the necessary time is much greater.

\subsubsection{No a priori spectral knowledge}\label{sec:noapriorispectralknowledge}
One might be concerned that the quality of RISS and CURISS depends on the availability of a good dictionary of the spectra $\mat{M}_{\mathrm{dictionary}}$. We show here that we obtain comparable results if we do not have access to a spectral dictionary.

\begin{figure}
    \centering
    \includegraphics[width=0.95\linewidth]{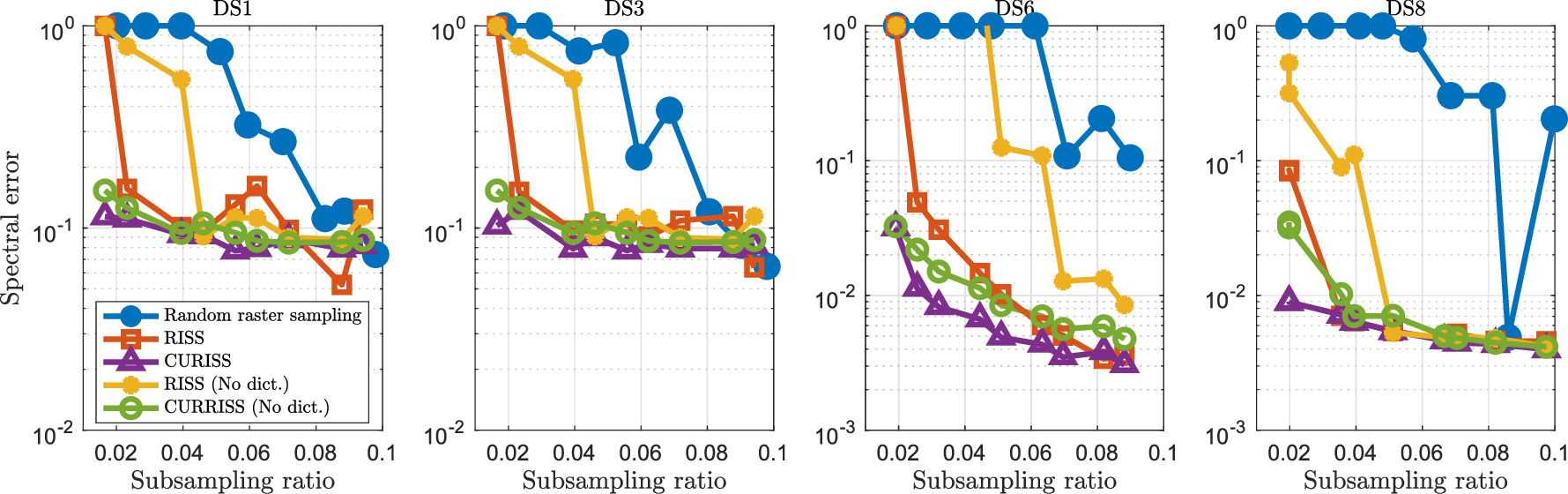}
    \caption{The spectral error (defined in Eq.~\ref{eq:spectral-error}) associated to each of the discussed algorithms, as well as RISS and CURISS without the use of a spectral dictionary $\mat{M}_{\mathrm{dictionary}}$ (no dict.). }
    \label{fig:spectral_error_no_dict}
\end{figure}

For each dataset, the quality of the completion for the algorithms with or without a dictionary is indistinguishable for subsampling ratios greater than 10\%. The results for ratios smaller than 10\% is shown in Figure~\ref{fig:spectral_error_no_dict}. For subsampling ratios smaller than 6\%, RISS without a dictionary suffers from considerably larger errors, while CURISS is more robust to this deficiency. 

\begin{figure}[t!]
    \centering
    \includegraphics[width=\linewidth]{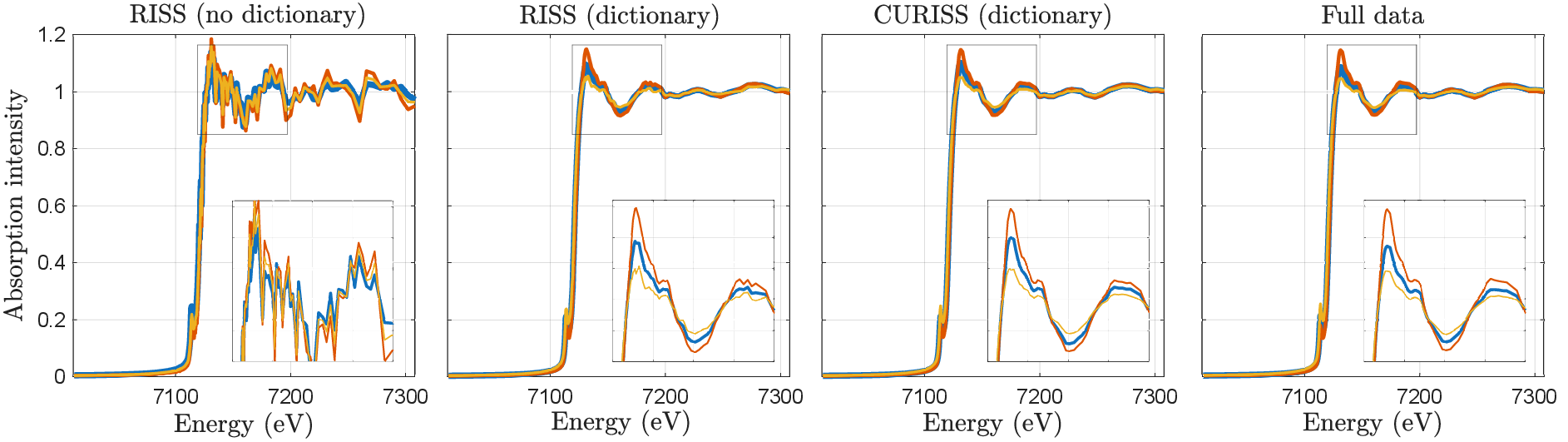}
    \caption{Spectra obtained from DS6 at 6\% subsampling for different sampling strategies. The right figure shows the spectra obtained from the full data. The spectral errors are from left-to-right 0.048, 0.006, and 0.004.}
    \label{fig:comp_spectra_dic_nodic}
\end{figure}

We also show a visualisation of the spectral errors in Figure~\ref{fig:spectral_error_no_dict} on obtained spectra. Figure~\ref{fig:comp_spectra_dic_nodic} shows the recovered spectra for DS6 at 6\% subsampling for RISS without a dictionary (spectral error 0.048), RISS with a dictionary (spectral error 0.006), and CURISS with a dictionary (spectral error 0.004), and shows the spectra obtained from the full data. 

\section{Conclusions}
We have proposed several data-driven approaches to subsampling during a spectro-microscopy experiment. The goal of these experiments is to determine bond length variations in materials with nanoscale spatial resolution. These experiments are generally limited by long acquisition times and radiation damage, as they require a raster XRF scan for various energy levels. Subsampling approaches to spectro-microscopy aim to mitigate this by omitting various measurements in a way that fits the raster acquisition of measurements, without changing the (spatial or spectral) resolution. Missing measurements are in-painted with low-rank matrix approximation methods using the available observations, e.g., matrix completion with LoopedASD, and matrix factorization with CUR. This is possible because of the low-dimensional nature of a spectro-microscopic dataset; the underlying dimension is equal to the number of spectroscopically different materials up to measurement noise. 

The question that might arise when considering subsampling approaches is to what degree can spatial subsampling be applied and what are the limits of the approach. Fundamentally, the approaches depend on the information content in the measurements being representative of the full measurement $A$, i.e. to recover $A$, the chemical states in $A$ must be represented in the sample set, otherwise it would not be recovered. Methods based on uniform randomly (raster) subsampling measurement, such as LoopedASD \cite{townsend2022undersampling}, suffer from the aforementioned drawback. We built on \cite{townsend2022undersampling} by making the random sampling data-driven. In the data-driven scheme, entries that contain more information are measured with higher probabilities. This data-driven approach allows us to measure very small amounts of data while still retaining much of the relevant information. In particular, we find that for the available datasets, data-driven approaches can reach subsampling levels of $2-5$\% whereas uniform random sampling would need $10-20$\% subsampling ratios.

We note especially the efficacy of the CUR-based approach. The interpretability of this approach, combined with the simplicity and speed of its corresponding completion, suggests potential for its application in different experiments with low-dimensionality. As far as the authors are aware, this is the first time that the CUR decomposition has been used as a tool for experiment design. Especially in the context of costly experiments, we believe this approach has great potential to speed up acquisition times and even improve resolution beyond current mechanical limits. 

Future work includes further investigation of adaptive approaches and stopping criteria. Once an initial completion was performed for an initial subsampling ratio, adaptively adding a new XRF scan, can be done via other criteria, such as restarting the leverage scores from each new completion. As for other stopping criteria, apart from testing the change between iterations via the error metrics, other approaches can be leveraged from numerical linear algebra, e.g., testing the singular values and numerical rank differences between iterations. In addition, our work can be generalized to the tensor framework, as done in \cite{townsend2025alternatingsteepestdescentmethods}, by using tensor generalizations for the notions of leverage scores and CUR decomposition. We leave it for future work as well.

\appendix 
\section{Algorithms}\label{sec:appendix-algorithms}
In this section, we provide details and write in pseudocode form our proposed algorithms, which we explain in Section \ref{sec:expapproaches}.

We begin with our algorithm to determine spectral and spatial importance distributions (Algorithm \ref{pseudoalgo:distributions}).
The spectral importance distribution is based on the leverage scores of a dictionary of ground-truth spectra. In practice, this is achieved by performing a QR factorization of $\mat{M}_{\text{dictionary}} = \mat{Q}\mat{R}$, followed by computing the row norms of $\mat{Q}$. Using these leverage scores as sampling weights, we then select $s_E$ out of the $n_E$ XRF scans to be measured. 
For the spatial importance distribution, we choose to apply Adaptive Randomized Pivoting (ARP) to the horizontally stacked XRF maps, $\mat{F} = \begin{bmatrix} \mat{F}_1 & \dots & \mat{F}_{s_E} \end{bmatrix}$. ARP is an algorithm that adapts leverage scores while selecting rows to sample. The process begins similarly to standard leverage score sampling by computing the matrix $\mat{U}_1$ of dominant left singular vectors of $\mat{F}$ and using the row norms of $\mat{U}_1$ as initial sampling probabilities. The first index $i_1$ is randomly selected according to these normalized leverage scores. Next, $\mat{U}_1$ is updated to remove the selected column $\mat{U}_1(:,i_1)$, resulting in a new matrix $\bar{\mat{U}}_1$. This is achieved via orthogonal projection implemented by Householder reflectors. The second index is then sampled based on the updated leverage scores of $\bar{\mat{U}}_1$. This process repeats until all indices are selected and $\mat{U}_1$ has been fully reduced. Note that this process is able to pick at most $r$ indices when applied to an $n\times r$ matrix. Thus, we cannot use it for the previous step, where from the $n_E\times \hat{S}$ dictionary $\mat{M}_{\text{dictionary}}$ we want to obtain importance scores to select $s_E\geq \hat{S}$ XRF scans.
\begin{algorithm}[h]
    \caption{Importance Sampling for Spectro-microscopy: determining spectral and spatial importance distribtions. \\ Input: subsampling ratio $p$, number of spatial rows $n_y$ and columns $n_X$ of the sample, number of energies $n_E$, and a dictionary of absorption spectra $\mat{M}_{\mathrm{dictionary}}$ (optional). \\
    Output: a small number of full XRF scans, a spectral importance distribution, and a spatial importance distribution}
\begin{algorithmic}[1]
\State Set $s_E = \left\lceil \frac{pn_Yn_E}{n_Y + n_E} \right\rceil$ and $s_R = \left\lfloor \frac{n_Y(pn_E-s_E)}{n_E-s_E} \right\rfloor$.
\State Use the dictionary $\mat{M}_{\mathrm{dictionary}}$ to estimate the leverage scores on the energies and define the corresponding spectral importance distribution. If a dictionary is not available, define the spectral importance distribution as uniform.
\State Sample $s_E$ energies $E_{s_1}, \dots, E_{s_{s_E}}$ from the spectral importance distribution.
\State Measure full raster scans at energies $E_{s_1}, \dots, E_{s_{s_E}}$. Name the corresponding XRF maps $\mat{F}_1, \dots, \mat{F}_{s_E}$.
\State Horizontally stack the XRF maps as $\mat{F}=\begin{bmatrix}
    \mat{F}_1 & \dots & \mat{F}_{S_E}
\end{bmatrix}$ and use $\mat{F}$ to determine a spatial importance distribution. \label{algline:spatial-importance-distribution}
\end{algorithmic}
\label{pseudoalgo:distributions}
\end{algorithm}

Next, we present pseudocode for the presented strategies. That is,  RISS (Subsection \ref{subsec:RISS}) in Algorithm \ref{pseudoalgo: RISS}, CURISS (Subsection \ref{subsec:CURISS}) in Algorithm \ref{pseudoalgo: CURISS}, and  ACURISS (Subsection \ref{subsec:adaptivity-theory}) in Algorithm \ref{pseudoalgo:adaptiveCURISS}.

\begin{algorithm}[h]
    \caption{RISS: Raster Importance Sampling for Spectro-microscopy. \\ Input: subsampling ratio $p$, number of spatial rows $n_y$ and columns $n_X$ of the sample, number of energies $n_E$, and a dictionary of absorption spectra $\mat{M}_{\mathrm{dictionary}}$ (optional). \\
    Output: an approximation to the dataset $\mat{D}$ using subsampling ratio $p$.}
\begin{algorithmic}[1]
\State Use Algorithm~\ref{pseudoalgo:distributions} to measure a small number of full XRF scans at energies $E_{s_1}, \dots, E_{s_{s_E}}$ and determine a spatial importance distribution. 
\For{all energies $E$ in $\{E_1,\dots,E_{n_E}\}\backslash\{E_{s_1}, \dots, E_{s_{s_E}}\}$}
\State Set the beam energy to $E$.
\State Sample $s_R$ spatial rows from the spatial importance distribution: $r_{s_1}^{(E)}, \dots, r_{s_R}^{(E)}$
\State Measure the sampled spatial rows: $r_{s_1}^{(E)}, \dots, r_{s_R}^{(E)}$
\EndFor
\State Complete the measured dataset using LoopedASD.
\end{algorithmic}
\label{pseudoalgo: RISS}
\end{algorithm}

\begin{algorithm}[t]
    \caption{CURISS: CUR Importance Sampling for spectro-microscopy\\ Input: subsampling ratio $p$, number of spatial rows $n_y$ and columns $n_X$ of the sample, number of energies $n_E$, and a dictionary of absorption spectra $\mat{M}_{\mathrm{dictionary}}$ (optional). \\
    Output: an approximation $\mat{\hat{D}}$ to the dataset $\mat{D}$ using subsampling ratio $p$.}
\begin{algorithmic}[1]
\State Use Algorithm~\ref{pseudoalgo:distributions} to measure a small number of full XRF scans at energies $E_{s_1}, \dots, E_{s_{s_E}}$ and determine a spatial importance distribution. 
\State Sample $s_R$ spatial rows from the spatial importance distribution.
\For{all energies $E$ in $\{E_1,\dots,E_{n_E}\}\backslash\{E_{s_1}, \dots, E_{s_{s_E}}\}$}
\State Set the beam energy to $E$.
\State Measure the sampled spatial rows
\EndFor
\State Complete the measured dataset using CUR.
\end{algorithmic}
\label{pseudoalgo: CURISS}
\end{algorithm}

\begin{algorithm}[t] \caption{ACURISS: Adaptive CURISS \\ Input: initial subsampling ratio $p_0$, $\eta_{\mat{D}}\geq 0$ and $\eta_{\mat{M}}\geq 0$ are tolerance parameters for the stopping criteria, number of spatial rows $n_Y$ and columns $n_X$ of the sample, number of energies $n_E$, and a dictionary of absorption spectra $\mat{M}_{\mathrm{dictionary}}$ (optional). \\
    Output: an approximation $\mat{\hat{D}}$ to the dataset $\mat{D}$, and its corresponding subsampling ratio $p$.}
\begin{algorithmic}[1]
\State Use CURISS (Algorithm \ref{pseudoalgo: CURISS}) to obtain $\mat{\hat{D}}_{0}$.
\State $i=0$
\While{$\|\mat{\Delta D}_{i}\| > \eta_{\mat{D}}$ or $\|\mat{\Delta M}_{i}\|>\eta_{\mat{M}}$}
\State Set to $0$ the leverage scores of the previously selected energies.
\State Sample one additional energy using the refined spectral importance distribution.
\State Set $i=i+1$.
\State Update the subsampling ratio $p_{i}$.
\State Update CUR with the additional XRF scan to obtain $\mat{\hat{D}}_{i}$.
\State Set $s_E = s_E + 1$.
\EndWhile
\State \Return $\mat{\hat{D}} = \mat{\hat{D}}_{i}$ and $p = p_{i}$.
\end{algorithmic}
\label{pseudoalgo:adaptiveCURISS}
\end{algorithm}

\section{Mathematical foundations}
In the following, we provide more details, mathematical background and implementation specifics about observations and statements in the main text.
\subsection{Further motivations on the use of a dictionary of true spectra}
\label{appx:Motivation-dictionary}
In Section \ref{subsec:spectral-importance} we have introduced the idea of using a dictionary of true spectra $\mat{M}_{\mathrm{dictionary}}$ to obtain approximate leverage scores of the matrix $\mat{D}$. This idea is based on the observation that, since $\mat{D}_{\mathrm{exact}} = \mat{M}\mat{T}$, the leverage scores of $\mat{D}_{\mathrm{exact}}$ are the same as the leverage scores of $\mat{M}$. We elaborate here.

Let $\mat{D}_{\mathrm{exact}} = \mat{U}_1 \mat{\Sigma}_1 \mat{V}_1^T$ be the (reduced) singular value decomposition of $\mat{D}_{\mathrm{exact}}$. That is, $\mat{U}_1 \in \mathbb{R}^{n_E\times S}$ and $\mat{V}_1 \in \mathbb{R}^{n_Xn_Y \times S}$ are matrices with orthonormal columns corresponding to the leading singular vectors of $\mat{D}_{\mathrm{exact}}$ and $\mat{\Sigma}_1 \in \mathbb{R}^{S\times S}$ is a diagonal matrix containing the first $S$ singular values of $\mat{D}_{\mathrm{exact}}$. Then, by definition, the leverage scores of $\mat{D}_{\mathrm{exact}}$ are
\begin{equation*}
    l_i^2 = \|\mat{U}_1(i,:)\|^2_2.
\end{equation*}
Now, let $\mat{M} = \mat{Q}\mat{R}$, with $\mat{Q} \in \mathbb{R}^{n_E\times S}, \mat{R}\in \mathbb{R}^{S\times S}$, and  $\mat{T}^T = \tilde{\mat{Q}}\tilde{\mat{R}}$, with $\tilde{\mat{Q}} \in \mathbb{R}^{n_Xn_Y\times S}, \tilde{\mat{R}}\in \mathbb{R}^{S\times S}$, be the QR-factorizations of $\mat{M}$ and $\mat{T}^T$. Then, we have
\begin{equation}
\label{eq:qr-expression-D-exact}
    \mat{D}_{\mathrm{exact}} = \mat{M}\mat{T} = \mat{Q}\mat{R}\tilde{\mat{R}}^T\tilde{\mat{Q}}^T.
\end{equation}
Considering the (full) singular value decomposition $\mat{R}\tilde{\mat{R}}^T = \hat{\mat{U}}\hat{\mat{\Sigma}}\hat{\mat{V}}^T$ and using it in \eqref{eq:qr-expression-D-exact}, we obtain
\begin{equation*}
    \mat{D}_{\mathrm{exact}} = \mat{Q}\hat{\mat{U}}\hat{\mat{\Sigma}}\hat{\mat{V}}^T\tilde{\mat{Q}}^T = \mat{U}_1 \mat{\Sigma}_1 \mat{V}_1^T,
\end{equation*}
and, thus, $\mat{U}_1 = \mat{Q}\hat{\mat{U}}$. Finally, since $\hat{\mat{U}}$ is a square orthogonal matrix, and the Euclidean norm is invariant under orthogonal transformations, we have
\begin{equation*}
    l_i^2 = \|\mat{U}_1(i,:)\|^2 = \|\mat{Q}(i,:)\hat{\mat{U}}\|^2 = \|\mat{Q}(i,:)\|^2,
\end{equation*}
i.e., the leverage scores of $\mat{D}_{\mathrm{exact}}$ are the same as the leverage scores of $\mat{M}$. 

When we turn our attention to $\mat{D}$, the situation is slightly diffeent. Since $\mat{D}$ is only approximately low-rank, its leverage scores will no longer coincide with those of $\mat{M}$. However, because we have $\mat{D} = \mat{M}\mat{T} + \mat{G}$, where $\mat{G}$ represents relatively small noise (i.e., $\|\mat{G}\|$ is small), we can reasonably expect the leverage scores of $\mat{D}$ to remain close to those of $\mat{M}$.

\subsection{Theoretical guarantees on the quality of the CUR decomposition}
In Section \ref{sec:CURintro} we provided a technical introduction to CUR decomposition. In this subsection, we provide some theoretical guarantees from the vast literature on CUR. In \cite{goreinov1997theory,boutsidis2014optimal}, it is stated that if $\mat{A}$ is exactly rank-$k$, then there exists a CUR decomposition that exactly recovers $\mat{A}$. In \cite{drineas2008relative}, the authors show that there exists a CUR of rank $k$ that provides, with high probability ($1-\delta$), the following.
\begin{equation}\label{eq:CUR_rel_error}
    \|\mat{A}-\mat{CU^\dagger R}\|_F \leq (1+\epsilon) \|\mat{A}-\mat{A}_{k}\|_F,
\end{equation}
where $\mat{A}_{k}$ is an optimal rank-$k$ approximation of $\mat{A}$, via the truncated SVD, and $\epsilon>0$ is an error parameter. The number of columns, $c$, that are selected from $\mat{A}$ using statistical leverage scores is $c=O(k^2 \log{(1/\delta)/\epsilon^2})$ exactly or $c=O(k\log{(k)} \log{(1/\delta)}/\epsilon^2)$ in expectation, and the number of rows, $r$, that are selected from $\mat{A}$ using statistical leverage scores is $r=O(c^2 \log{(1/\delta)}/\epsilon^2))$ exactly or $r=O(c\log{(c)} \log{(1/\delta)}/\epsilon^2)$ respectively. In \cite{boutsidis2014optimal}, a deterministic (not probabilistic) bound similar to Eq. \eqref{eq:CUR_rel_error} is provided with the number of selected rows and columns is reduced to $O(k/\epsilon)$.

\begin{acknowledgements}
We thank Taejun Park, Nathaniel Pritchard, Yuji Nakatsukasa, Johnathan M. Bulled, Julia Parker, and Miguel Gomez Gonzalez for useful discussions.
\end{acknowledgements}

\begin{funding}
This work was funded by the Ada Lovelace Centre
Programme at the Scientific Computing Department, STFC.
\end{funding}

\ConflictsOfInterest{The authors declare no conflicts of interest.
}

\DataAvailability{Data used to produce the ﬁgures and tables in this manuscript are available for downloading at the following open-access Zenodo repository: 10.5281/zenodo.18470992 \cite{Meier_Subsampling_imaging_spectro_2026}.}

\bibliography{iucr} 

\end{document}